\documentclass[usenatbib]{mnras}
\usepackage{graphicx}
\usepackage{float}
\usepackage{amsmath}
\usepackage{amssymb}
\usepackage{textcomp}
\usepackage{multirow}  
\usepackage[caption=false]{subfig}
\usepackage{color}
\usepackage{gensymb}   
\usepackage{hyperref} 
\usepackage{xspace} 
\usepackage{afterpage} 

\newcommand{\comment}[1]{\ignorespaces}
\newcommand{\HI}{\textsc{Hi}\xspace}

\makeatletter
\newcommand\footnoteref[1]{\protected@xdef\@thefnmark{\ref{#1}}\@footnotemark}
\makeatother

\title[Strangulation in Low-Mass Hosts: DDO 113]{The Case for Strangulation in Low-Mass Hosts: DDO 113}

\author[C. T. Garling et al.]{Christopher T. Garling,$^{1}$\thanks{E-mail: garling.14@osu.edu}
Annika H. G. Peter,$^2$
Christopher S. Kochanek,$^1$ \newauthor
David J. Sand,$^3$
Denija Crnojevi\'c$^4$
\\
$^1$CCAPP and Department of Astronomy, The Ohio State University, 140 W. 18th Ave., Columbus, OH 43210, USA \\
$^2$CCAPP, Department of Physics, and Department of Astronomy, The Ohio State University, 191 W. Woodruff Ave., Columbus, OH 43210, USA \\
$^3$Department of Astronomy and Steward Observatory, University of Arizona, 933 N. Cherry Avenue, Tucson, AZ 85719, USA \\
$^4$University of Tampa, 401 West Kennedy Boulevard, Tampa, FL 33606, USA \\
}

\date{Accepted XXX. Received YYY; in original form ZZZ}

\pubyear{2019}

\begin{document}
\label{firstpage}
\pagerange{\pageref{firstpage}--\pageref{lastpage}}
\maketitle

\begin{abstract}
We investigate the case for environmental quenching of the Fornax-mass satellite DDO 113, which lies only 9 kpc in projection from its host, the Large-Magellanic-Cloud-mass galaxy NGC 4214. DDO 113 was quenched about 1 Gyr ago and is virtually gas-free, while analogs in the field are predominantly star-forming and gas-rich. We use deep imaging obtained with the Large Binocular Telescope to show that DDO 113 exhibits no evidence of tidal disruption to a surface brightness of $\mu_V\sim29$ mag $\text{arcsec}^{-2}$, based on both unresolved emission and resolved stars. Mass-analogs of DDO 113 in Illustris-1 with similar hosts, small projected separations, and no significant tidal stripping first fell into their host halo 2--6 Gyr ago, showing that tidal features (or lack thereof) can be used to constrain infall times in systems where there are few other constraints on the orbit of the satellite.  With the infall time setting the clock for environmental quenching mechanisms, we investigate the plausibility of several such mechanisms. We find that strangulation, the cessation of cold gas inflows, is likely the dominant quenching mechanism for DDO 113, requiring a time-averaged mass-loading factor of $\eta=6-11$ for star-formation-driven outflows that is consistent with theoretical and observational constraints. Motivated by recent numerical work, we connect DDO 113's strangulation to the presence of a cool circumgalactic medium (CGM) around NGC 4214. This discovery shows that the CGM of low-mass galaxies can affect their satellites significantly and motivates further work on understanding the baryon cycle in low-mass galaxies.
\end{abstract}


\section{Introduction}
Why do galaxies form stars?  Why do they stop?  For large galaxies ($M_* \geq 10^8 \, M_{\odot}$), we can use wide-field surveys to study star formation across cosmic time in varied environments to reveal statistical trends. For small galaxies, detailed studies of star formation have been limited to the Local Group and a handful of Local Volume galaxies. The focus of this work is how to fill this gap and learn more about the end of star formation in low-mass galaxies outside the Local Group. \par

A key question of galaxy evolution is why galaxies go from blue and star-forming to red and quenched. For large galaxies ($M_* \geq 10^8 \, M_\odot$), photometry can be aggregated with other observational data to show that environment must play a significant role in quenching star formation \citep[e.g.,][]{Kawinwanichakij2017,Darvish2018,Lin2019}.  Large, red galaxies have long been known to cluster in denser environments \citep[see, e.g.,][]{Dressler1980,Butcher1984,Hogg2003,Kauffmann2004,Cooper2006,Haines2007}.  From large samples of galaxies, the combination of lensing, two-point correlation statistics, and the time-evolution of the galaxy stellar mass function indicate that the host halo mass drives quenching \citep{Woo2013,Zu2016,Moutard2018,Kawinwanichakij2017,Papovich2018}.  Based on the quenching timescales \citep{Wetzel2012,wheeler2014} and stellar and gas-phase metallicities \citep{Peng2015,Trussler2018,Maier2019} inferred for large ensembles of galaxies, it seems likely that a combination of starvation (the cessation of gas accretion onto galaxies) and outflows from star formation (sometimes called ``overconsumption") play a significant role in ending star formation in dense environments \citep{Larson1980,McGee2014,Peng2015,Maier2019}.  On the other hand, the ``slow then fast" quenching timescales found by \citet{Wetzel2012} indicate that the ``fast" end phase of quenching may be dominated by the removal of cool star-forming gas by ram-pressure stripping \citep{Gunn1972}, a claim supported by studies of cluster gas density measurements and simulations \citep{Lotz2018,Roberts2019,Tremmel2019,Tonnesen2019}.  

Environment seems to overwhelmingly drive quenching for the lowest mass galaxies typically observed in large numbers in wide-field surveys ($10^8 \, M_\odot < M_* < 10^{10} \, M_\odot$), as almost no field galaxies in this mass range are quenched in low-density environments \citep{Haines2007,Geha2012,Kawinwanichakij2017}. The prevalence of quenched massive galaxies in the field suggests that internal processes like AGN feedback and virial shock heating plays a key role in terminating star formation in big galaxies \citep{Forbes2016,DiMatteo2005,Birnboim2003,Wang2008,Keres2009}.  Gravitational interactions between hosts and satellites, and between satellites, appear to play a minor role in quenching, although they can be important for morphological transformation \citep{Kazantzidis2011,Deason2014}.

By contrast, for dwarf galaxies below $M_* < 10^8 M_\odot$, most work on quenching has focused on the unique environment of the Milky Way, where individual galaxies can be studied in detail using a variety of measurements unavailable for more distant galaxies \citep{Mayer2006,Grcevich2009,nichols2011,Rocha2012,Gatto2013,slater2014,Weisz2015,Wetzel2015,fillingham2016,Fillingham2019}.  Many of these studies exploit the six-dimensional phase-space information measured for each galaxy, the accuracy of which has greatly improved with GAIA-DR2 \citep{GaiaCollaboration2018}. The orbital information can be used to estimate satellite infall times, which set the clock for how fast quenching processes must act, and to estimate the history of the tidal and ram-pressure forces acting on the satellites. These can be combined with star-formation histories (SFH) derived from studying the resolved stellar populations to show how star formation tracks orbits \citep{Dolphin2002a,Weisz2014b}.  Additionally, and unique to the Milky Way, all satellites but the Magellanic Clouds are devoid of neutral hydrogen \citep{Grcevich2009,Spekkens2014} and quenched.  

Together, these detailed studies of individual Milky Way satellites reveal that there is a split in the quenching pathways of ``ultra-faint" ($M_* < 10^5 \, M_\odot$) and ``classical" ($M_* > 10^5 \, M_\odot$) dwarfs, with the former being almost certainly quenched by reionization \citep{benson2002,Brown2014,Weisz2014b,RodriguezWimberly2019}, and the latter by environment. Various studies implicate ram-pressure stripping on account of the Milky Way's hot gas halo \citep{Gatto2013,slater2014,fillingham2016}.

However, these studies are specific to one environment: the Milky Way.  Advancing our understanding of quenching at low masses requires a diverse sample of dwarfs across a broader range of environments, as studies of more massive galaxies show that both satellite mass and environmental density matter for quenching mechanisms and timescales. In this study, we propose an approach that uses some elements of these Milky-Way-type studies of individual dwarf galaxies but with the less fine-grained data typical of extragalactic observations---in particular, the lack of six-dimensional phase-space information. As with Milky Way studies, we select a collection of analogs to observed host-satellite systems from cosmological simulations, building probability distribution functions for the infall times and orbital parameters of observed satellites based on the properties of the simulated systems. Producing tight constraints on orbits and quenching mechanisms using this method requires identification of observable quantities that can be used to select analogs. In practice this amounts to identifying observables that correlate with infall time and orbital parameters, such that applying constraints based on these observables produces a final sample of analog systems whose properties accurately reflect those of the observed system. \par

Here we will show that the tidal disruption state and projected distance can be used to estimate the infall time and orbital parameters of extragalactic satellites, even when no kinematic data are available. We will show that these results can be combined with analytic prescriptions for quenching mechanisms and new insight into the importance of the cool/warm circumgalactic medium (CGM) of low-mass central galaxies to provide strong constraints on quenching pathways for low-mass dwarf galaxies in a variety of environments.

Our approach is especially applicable to galaxies in the Local Volume, where the ability to resolve stellar populations allows for the detection of extremely low surface brightness tidal debris.  As a test case, we consider DDO~113, a Fornax-mass ($M_* = 1.8 \times 10^{7} \, M_{\odot}$) satellite galaxy only $\sim9$ kpc in projection from its host, the Large-Magellanic-Cloud-mass galaxy NGC~4214 \citep[$M_* = 3.29 \times 10^{9} \, M_{\odot}$;][]{Weisz2011}. DDO 113 is the only known satellite of NGC 4214, and resolved CMD modelling shows that it ceased star formation about 1 Gyr ago \citep{Weisz2011}, while interferometric 21-cm \HI data shows that it is virtually gas-free \citep{Walter2008,Ott2012,Hunter2012}. This is unexpected, as DDO 113 analogs in the field are predominantly star-forming \citep{Geha2012,Fillingham2018} and gas-rich \citep{Papastergis2012}, suggesting that environmental quenching may be responsible. DDO 113 lies at a distance of $2.95 \pm 0.083$ Mpc \citep{Dalcanton2009}, which allows us to leverage resolved red giant branch stars as well as unresolved light in our deep imaging with the Large Binocular Cameras \citep[LBC;][]{Ragazzoni2006,Speziali2008} on the Large Binocular Telescope \citep[LBT;][]{Hill2010a}. \par 

This is the first of several systems we are studying with our survey of satellites of nearby star-forming galaxies with the LBT (LBT-SONG) in concert with the MADCASH surveys \citep{Carlin2019,Hargis2019} to understand quenching as a function of environment and satellite/host stellar mass. In \S\ref{section:data} and \S\ref{section:TidalDisruption}, we present and use LBT data to show that DDO~113 has no tidal features down to a detection limit of $\mu_V = 29 \ \text{mag} \ \text{arcsec}^{-2}$. We use the observational constraints derived from our imaging data to estimate DDO 113's infall time and orbital parameters in \S \ref{section:illustris}, and use these estimates to investigate the effectiveness of different quenching mechanisms in \S \ref{section:quenching}. We discuss the results in \S \ref{section:conclusion}.\par

\section{Observations and Data Reduction} \label{section:data}

Our observations were originally obtained for the search for failed supernovae described in \cite{Kochanek2008}, with recent results in \cite{Adams2017a}. These observations used the blue and red channels of the Large Binocular Camera \citep[LBC;][]{Ragazzoni2006,Speziali2008} on the LBT \citep{Hill2010a}. The LBT has two 8.4m primary mirrors and the LBC consists of two prime focus cameras, with one optimized for blue light and one for red. Each camera uses four 2048 $\times$ 4608 pixel CCDs with a scale of $0\farcs225$ per pixel and a field of view of $23^{\prime} \times 23^{\prime}$, corresponding to an area of (20 kpc)$^2$ at the distance of DDO 113.\par

Our observations were taken as part of a monitoring program, with the earliest data for this system obtained in 2009 and the latest taken in 2017. Typically two epochs of data were obtained each year. Each epoch consisted of three two-minute \emph{R} band exposures on LBC red paired with single two-minute \emph{U}, \emph{B}, and \emph{V} band exposures on LBC blue. Seeing varied from $0\farcs8$ to $1\farcs5$, but we only used images with seeing better than $1\farcs1$. We used about 13 images (26 minutes of exposure) in each of \emph{U}, \emph{B}, and \emph{V} bands and 24 images (48 minutes of exposure) in \emph{R} band. Our data reduction, photometry, calibration, and artificial star tests draw inspiration from \cite{Sand2009} and \cite{Garling2018}. This analysis pipeline will be used for an upcoming satellite galaxy survey using the full data set of \cite{Kochanek2008}. \par

\subsection{Data Reduction}\label{subsection:reduction}
Raw images were processed with overscan correction, bias subtraction, and flat fielding with the \textsc{iraf mscred} package as described in \cite{Gerke2015}. Astrometric calibration was performed in two steps; we used a local installation of the \texttt{astrometry.net} code base \citep{Lang2010} with custom astrometric reference files built from SDSS-DR13 \citep{Alam2015} to produce an initial astrometric solution, which we then analyzed with \texttt{SCAMP} \citep{Bertin2006}. Due to the LBT's low focal ratio of $f/1.14$, there is significant optical distortion across the field of view, which we correct with \texttt{SCAMP} by fitting a third-degree polynomial. \texttt{SCAMP} then performs both internal and external astrometric calibrations, cross-matching the astrometry of our input images to standard deviations $\sim0\farcs1$ and calibrating to GAIA-DR2 \citep{GaiaCollaboration2018} sources with similar precision. \par

Following this astrometric calibration, we co-add our individual exposures using \texttt{SWarp} \citep{Bertin2002}. For co-addition, we used the clipped-mean algorithm from \cite{Gruen2014}, which can reject single-epoch image artifacts, resulting in cleaner final image co-adds than an inverse-variance weighted mean algorithm. The final co-added images have mean seeing of $\sim1\farcs1$. \par

\subsection{Photometry and Calibration} \label{subsection:photometry}
We performed PSF fitting photometry using the \textsc{daophot}, \textsc{allstar}, and \textsc{allframe} packages \citep{Stetson1987,Stetson1994}. We ran two passes of \textsc{allstar}: the first on the image, and the second on the image with detections from the first pass removed, allowing us to find faint stars missed on the first pass. We compute pixel coordinate transformations between filters by feeding the final \textsc{allstar} catalogs to the \textsc{daomatch/daomaster} packages \citep{Stetson1993}. Finally, we performed forced photometry simultaneously in all bands for sources detected in at least two filters by \textsc{allstar} using \textsc{allframe} \citep{Stetson1994}, increasing our photometric depth. As we wish to retain only well-measured stars, we first trim the \textsc{allframe} catalog by eliminating stars with photometric errors greater than 0.4 mag or $\chi$, a measure of the goodness-of-fit, greater than 2. Our primary statistic for separating stars and galaxies is the \textsc{daophot} sharpness parameter, which has been shown to work well \citep[e.g., Figure 4 in][]{Annunziatella2013} -- we require sources to have absolute values of sharpness less than 2 in order to be classified as stars. \par

After creating our final photometric catalog, we calibrate our photometry by bootstrapping onto SDSS-DR13 \citep{Alam2015} and correcting for Galactic extinction. We used relations from \cite{Jordi2006} to convert SDSS magnitudes to  \emph{U}, \emph{B}, \emph{V}, and \emph{R} with full error propagation. We match our photometric catalog to the SDSS-DR13 stars catalog, and use the stars in common to fit zero points and color terms for all bands simultaneously using an extended version of the maximum likelihood method described in \cite{Boettcher2013} that accounts for covariances between the zero points and color terms for our four bands. We correct for color terms in $(U_{\text{SDSS}}-U_{\text{LBT}})$ versus $(U_{\text{LBT}}-B_{\text{LBT}})$, $(B_{\text{SDSS}}-B_{\text{LBT}})$ versus $(B_{\text{LBT}}-R_{\text{LBT}})$, $(V_{\text{SDSS}}-V_{\text{LBT}})$ versus $(B_{\text{LBT}}-V_{\text{LBT}})$, and $(R_{\text{SDSS}}-R_{\text{LBT}})$ versus $(B_{\text{LBT}}-R_{\text{LBT}})$. Mean calibration errors are $0.03-0.05$ mag, determined by 10,000 iterations of bootstrap resampling.\par

Reported magnitudes are corrected for Galactic extinction with $E \, (B-V)$ values obtained by interpolating the dust maps from \cite{Schlegel1998} with the updated scaling from \cite{Schlafly2011}. The average extinction in the direction of DDO 113 is $E \, (B-V)=0.017$. Hereafter, all quoted magnitudes have been corrected for this extinction.

\begin{figure}
	\centering
        \includegraphics[width=0.45\textwidth,page=1]{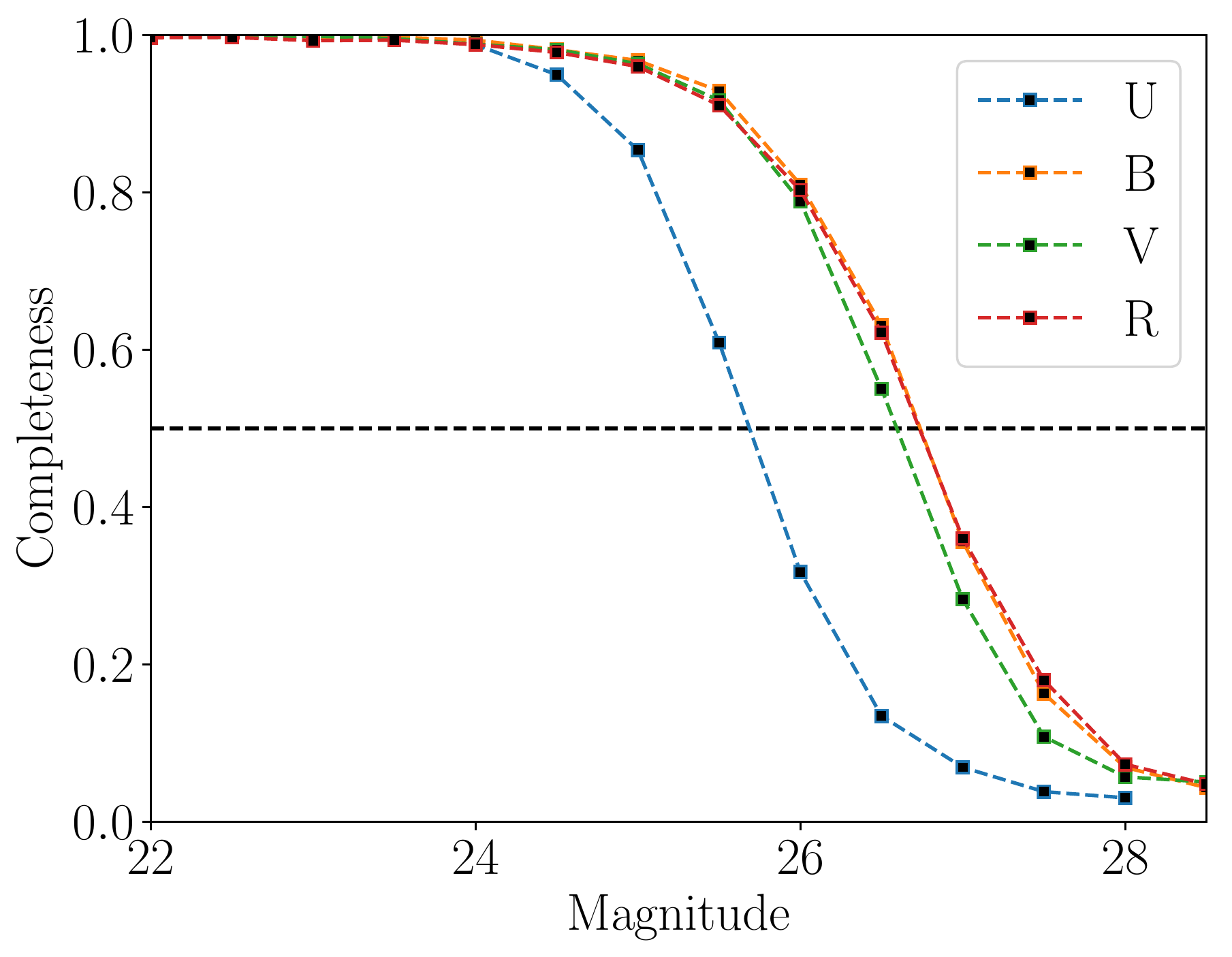}
	\caption{Point-source completeness curves in all bands derived from our artificial star tests. The same quality cuts have been applied as were used for our final photometric catalog.} 
	\label{completeness_curve}
\end{figure}

\subsection{Artificial Star Tests} \label{subsection:artificialstartests}
To measure our photometric errors and completeness as a function of magnitude and color, we perform artificial star tests in all bands on our final co-adds, excluding the central chip where NGC 4214 occupies a significant portion of the total area. While DDO 113 lies solely on a single chip, we inject and analyze mock DDO 113 analogs on three chips in \S \ref{subsection:galfit} to estimate errors in our morphological parameter measurements -- for this application, we need to ensure that our photometric depths on all chips are comparable, which we are able to confirm. \par 

We create a mock catalog of artificial stars on a spatial grid with a $\sim15^{\prime\prime}$ spacing. Magnitudes in \emph{U} band are drawn from a uniform distribution from 22 to 28.5 mag, and magnitudes in the other bands are determined by drawing colors from uniform distributions from $-1$ to $1$ in $U-B$, $B-V$, and $V-R$. These ranges were chosen to represent a reasonable distribution of stellar colors. We use the \textsc{addstar} routine from \textsc{daophot} \citep{Stetson1987,Stetson1994} to insert the stars, which makes full use of the best-fit analytic form of the PSF, its empirical correction table, and the spatial variation of the PSF \citep[see][]{Stetson1992}. We create 10 artificial images per chip per band, each with 2,000 artificial stars for a total of 80,000 artificial stars. The artificial images are processed with the same photometric pipeline as our science images and we apply the same cuts in $\chi$, sharpness, and magnitude error to the resulting photometric catalog. Our completeness curves, averaged over the three chips considered, are shown in Figure \ref{completeness_curve}. We achieve 90\% (50\%) completeness at $24.8$ ($25.8$) magnitude in \emph{U} band and $25.8$ ($26.8$) magnitude in \emph{B}, \emph{V}, and \emph{R} bands. We find that our $5 \, \sigma$ point source measurement limits are roughly equivalent to our 90\% completeness limits. No significant differences in photometric depth or quality are observed between the three chips.

\section{Searching for Tidal Disruption in DDO 113} \label{section:TidalDisruption}
Given DDO 113's small projected distance ($\sim 9$ kpc) from NGC 4214, one scenario for the quenching of its star formation is a recent close encounter with NGC 4214 that tidally stripped baryonic material from the dwarf. To search for observational signs of tidal disruption, we use both \textsc{galfitm} (\citealt{Haeusler2013}; based on \textsc{galfit} version 3 from \citealt{Peng2010}), and elliptical isophotes to look for distortions in the diffuse light profile. We judge the sensitivity of these approaches by adding DDO 113 analogs to the images, both with and without distortions, and analyzing them in the same manner as the data. These methods primarily utilize unresolved light from faint stars. We additionally search for stellar streams in resolved stars and use image simulations to estimate our sensitivity.\par

\subsection{GALFITM Modelling} \label{subsection:galfit}
The standard version of \textsc{galfit} \citep{Peng2010} fits features in individual images using combinations of analytic models convolved with the estimated image PSF. \textsc{galfitm} \citep{Haeusler2013} inherits the flexibility of \textsc{galfit} while adding the ability to fit multiple bands simultaneously, with an adjustable amount of model flexibility between bands. \par

\begin{figure*}
  \centering
  \includegraphics[width=.3\linewidth]{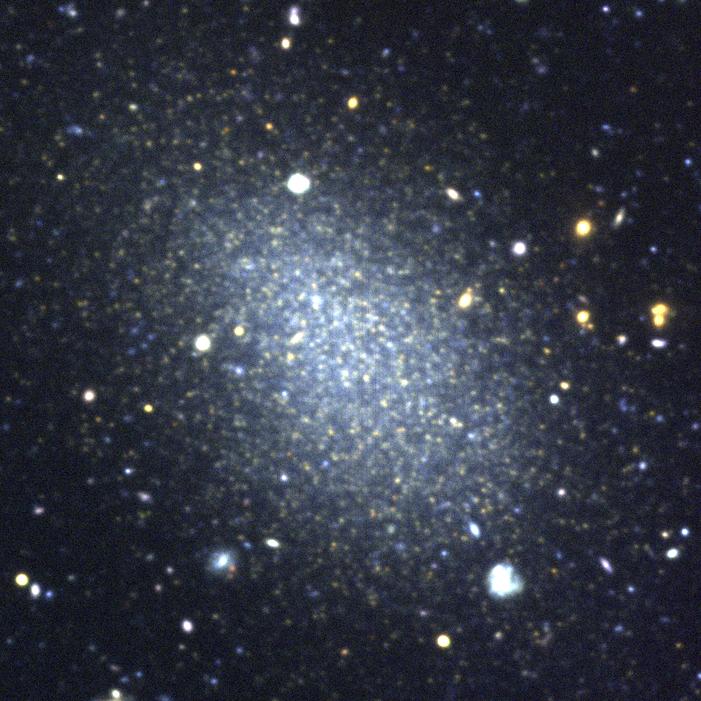}
  \includegraphics[width=.3\linewidth]{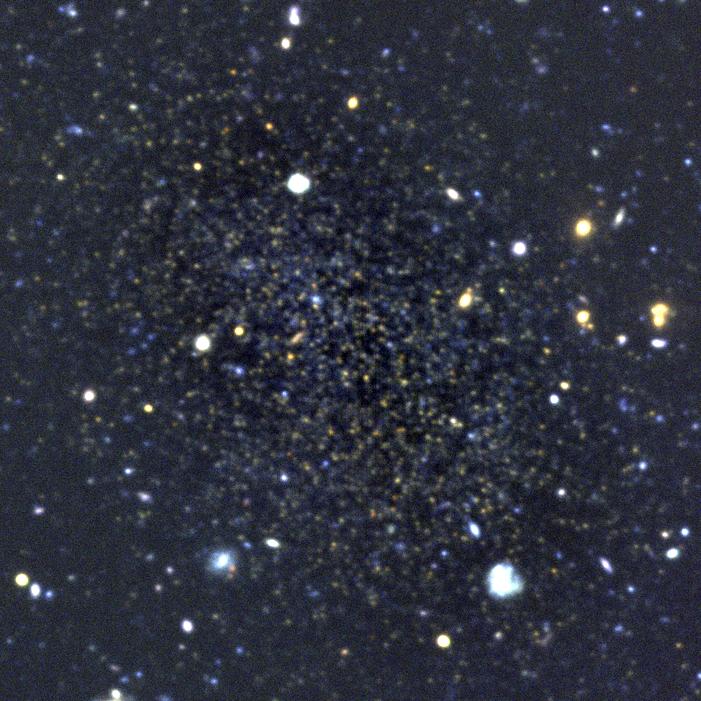}
  \includegraphics[width=.3\linewidth]{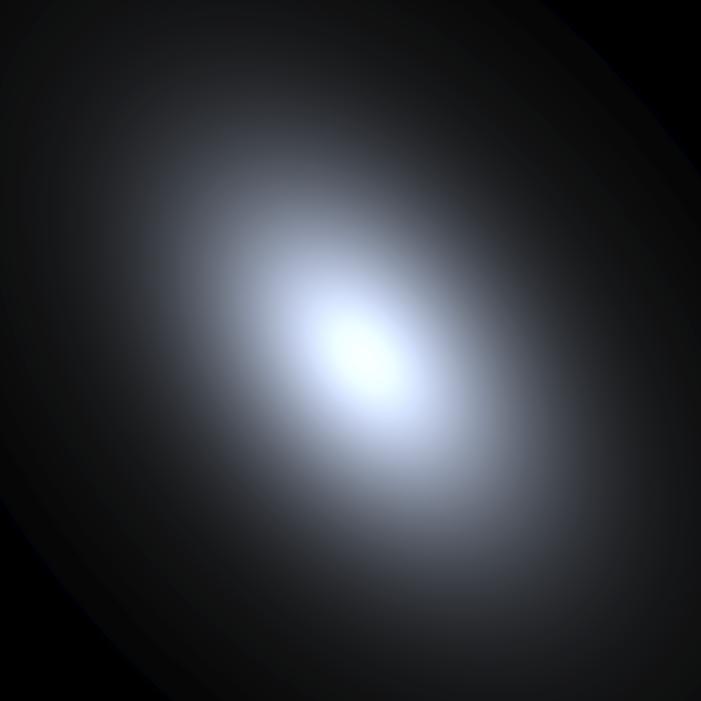}
  \caption{\emph{Left:} A composite RGB (\emph{R}, \emph{V}, \emph{B} bands) color image of DDO 113. \emph{Center:} The residuals after subtracting the best-fit S\`ersic model. \emph{Right:} The best-fit \textsc{galfitm} S\`ersic model. The color scale of the model image has been adjusted to match the science images and the contrast of the subtracted image has been increased to make the residuals visible.}
  \label{galfitmodels}
\end{figure*}

We use \textsc{galfitm} to fit an axisymmetric 2D elliptical S\`ersic profile to the diffuse light profile of DDO 113 in all bands and then search for asymmetries in the fit residuals which could indicate tidal disruption. Table \ref{Table:galfitresults} gives the results from fitting each band separately using \textsc{galfit} and fitting all bands simultaneously with a single set of profile parameters with \textsc{galfitm}. The latter approach is analogous to forced photometry, whereby flux is extracted from an image using an identical model in all bands, and is an established technique for precise magnitude and color measurements in stellar photometry (e.g., \textsc{allframe}; \citealt{Stetson1993}). The two sets of results are mutually consistent and reinforce DDO 113's similarity to Fornax, as both are elliptical ($q\approx$ 0.6--0.7) with low S\`ersic indices ($n\approx$ 0.6--0.7; \citealt{Munoz2018}) and comparable stellar masses ($M_*\approx2 \times 10^7 \, M_{\odot}$; \citealt{Weisz2014a}). Our \emph{B} band absolute magnitude of $-11.69$ is consistent to within the 0.1 mag error with the measurement of \cite{Karachentsev2013}. Figure \ref{galfitmodels} shows color images of DDO 113, the fit residuals, and the \textsc{galfitm} model. We see no significant asymmetries or substructures in the residuals, indicating a lack of photometric features that could be attributed to tidal disruption. \par

\begin{table}
  \caption{DDO 113 fitting results. Errors are mean squared errors based on measurements of nine DDO 113 analogs produced using the methods described in \S \ref{subsection:galfit}. } 
  \scalebox{0.8}{
    \hspace*{-3.5em}
    \begin{tabular}{c c c c c c}
      \hline
      \hline
      Parameter & U & B & V & R & Error  \\
      \hline
      \multicolumn{6}{c}{\textsc{galfit}} \\
      \hline
      R.A. & 12\textsuperscript{h}14\textsuperscript{m}58.3\textsuperscript{s} & 12\textsuperscript{h}14\textsuperscript{m}58.3\textsuperscript{s} & 12\textsuperscript{h}14\textsuperscript{m}58.3\textsuperscript{s} & 12\textsuperscript{h}14\textsuperscript{m}58.3\textsuperscript{s} & 0.1\textsuperscript{s}\\
      Decl. & $+36\degree13^{\prime}08^{\prime\prime}$ & $+36\degree13^{\prime}06^{\prime\prime}$ & $+36\degree13^{\prime}07^{\prime\prime}$ & $+36\degree13^{\prime}07^{\prime\prime}$ & $2^{\prime\prime}$ \\
      m\textsubscript{0} & 15.57 & 15.69 & 15.17 & 15.04 & 0.10\\
      $\mu_0$ & 24.74 & 24.79 & 24.33 & 24.22 & 0.12\\
      $\mu_e$ & 25.77 & 25.80 & 25.35 & 25.31 & 0.12\\
      R\textsubscript{e} & $43\farcs5$ & $42\farcs9$ & $43\farcs6$ & $44\farcs6$ & $1\farcs5$\\
        & 622 pc & 614 pc & 624 pc & 638 pc & 21 pc\\
      n & 0.63 & 0.62 & 0.63 & 0.66 & 0.02\\
      q & 0.64 & 0.61 & 0.63 & 0.64 & 0.05\\
      PA & $42\degree$ & $40\degree$ & $40\degree$ & $40\degree$ & $2\degree$\\
      \hline
      \multicolumn{6}{c}{\textsc{galfitm}} \\
      \hline
      R.A. & 12\textsuperscript{h}14\textsuperscript{m}58.3\textsuperscript{s} & -- & -- & -- & 0.1\textsuperscript{s}\\
      Decl. & $+36\degree13^{\prime}07^{\prime\prime}$ & -- & -- & -- & $2^{\prime\prime}$ \\
      m\textsubscript{0} & 15.62 & 15.66 & 15.16 & 15.09 & 0.10\\
      $\mu_0$ & 24.76 & 24.80 & 24.30 & 24.23 & 0.12\\
      $\mu_e$ & 25.81 & 25.85 & 25.35 & 25.28 & 0.12\\
      R\textsubscript{e} & $43\farcs6$ & -- & -- & -- & $1\farcs5$\\
       & 624 pc & -- & -- & -- & 21 pc\\
      n & 0.64 & -- & -- & -- & 0.02\\
      q & 0.63 & -- & -- & -- & 0.05\\
      PA & $40\degree$ & -- & -- & -- & $2\degree$\\
      \hline
      
    \end{tabular}
  } 
  \label{Table:galfitresults}
\end{table}

To judge the significance and uncertainties of the \textsc{galfit} results, we produce analogs of DDO 113 star by star. The stellar positions are sampled from 2D S\`ersic profiles with the morphological parameters measured by \textsc{galfitm}, while the stellar magnitudes are sampled from PARSEC 1.2S isochrones \citep{Aringer2009,Bressan2012,Chen2014} including the thermally-pulsating asymptotic giant branch and other improvements from \cite{Marigo2017}. We use three different simple stellar populations: 10 Gyr, $\text{[Fe/H]}=-2$, 5 Gyr, $\text{[Fe/H]}=-1$, and 1 Gyr, $\text{[Fe/H]}=0$. We use the \cite{Chabrier2001} lognormal initial mass function (IMF) and add the Galactic extinction derived in \S \ref{subsection:photometry} to the stellar magnitudes. Stars are added to the mock galaxies iteratively until a total of $M_V=-12.19$ is reached, corresponding to $m_V=15.16$ as found by \textsc{galfitm}. When all star positions and magnitudes have been generated, they are injected into the LBT images by \textsc{addstar} as in \S \ref{subsection:artificialstartests}.\par

\begin{figure*}
  \centering
  \includegraphics[width=.3\linewidth]{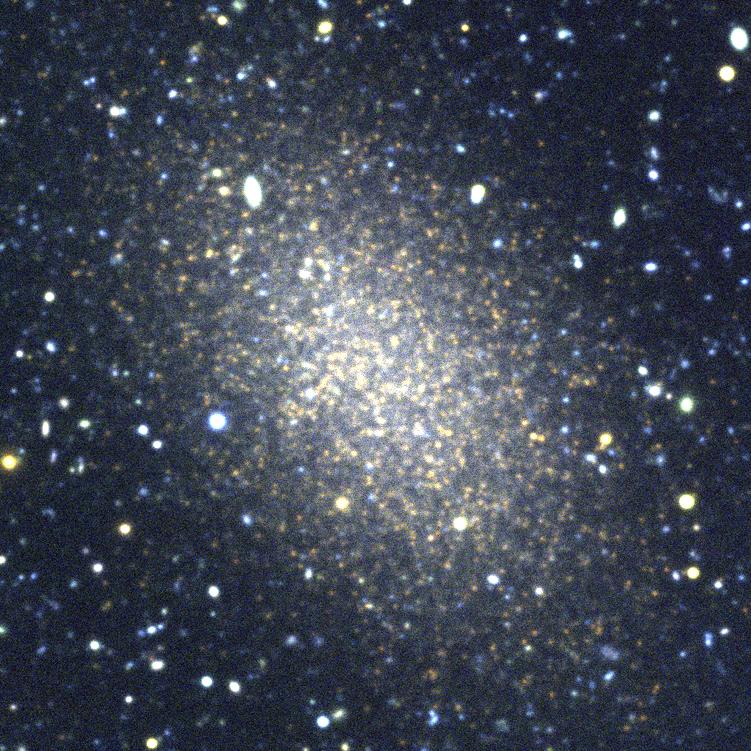}
  \includegraphics[width=.3\linewidth]{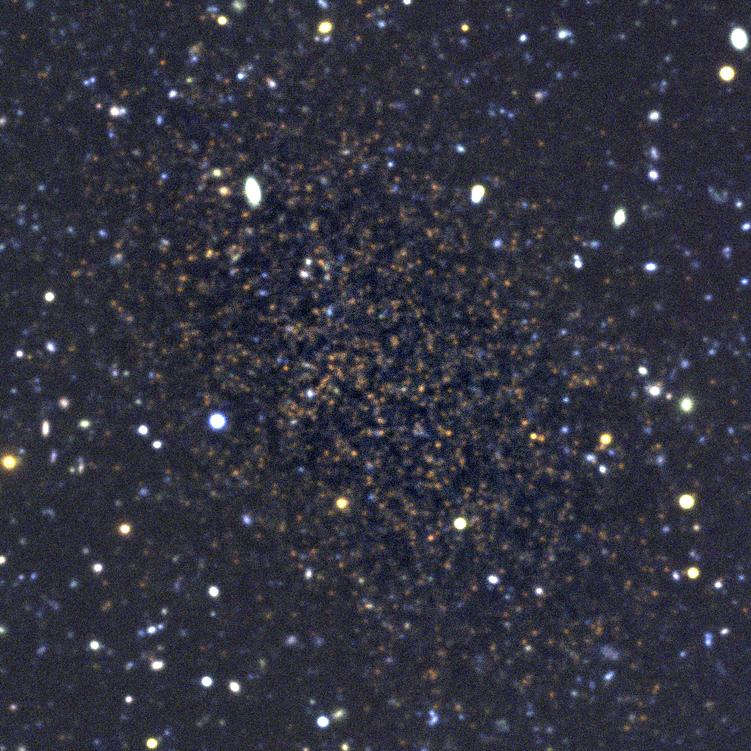}
  \includegraphics[width=.3\linewidth]{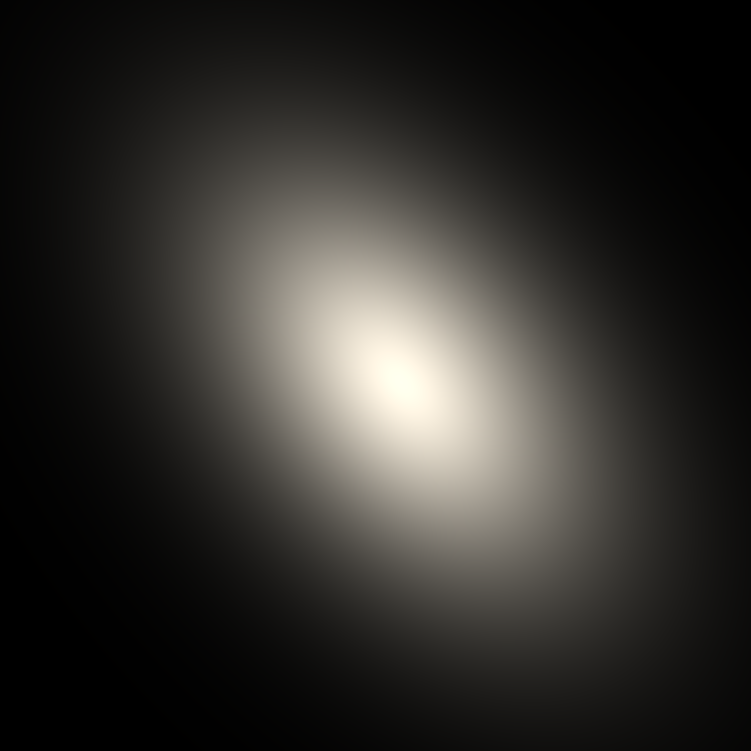}
  \caption{As in Figure \ref{galfitmodels} but for a simulated DDO 113 generated with a 5 Gyr, [Fe/H]$=-1$ isochrone.}
  \label{galfitmodels_fake5gyr}
\end{figure*}

\begin{figure*}
  \centering
  \includegraphics[width=.3\linewidth]{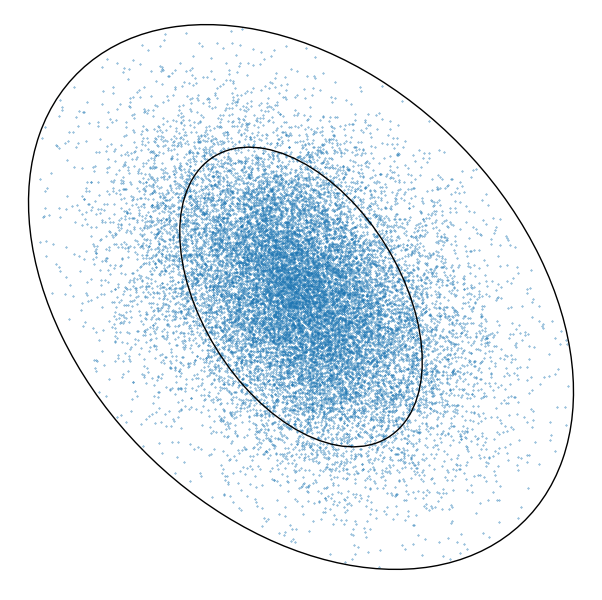}
  \includegraphics[width=.3\linewidth]{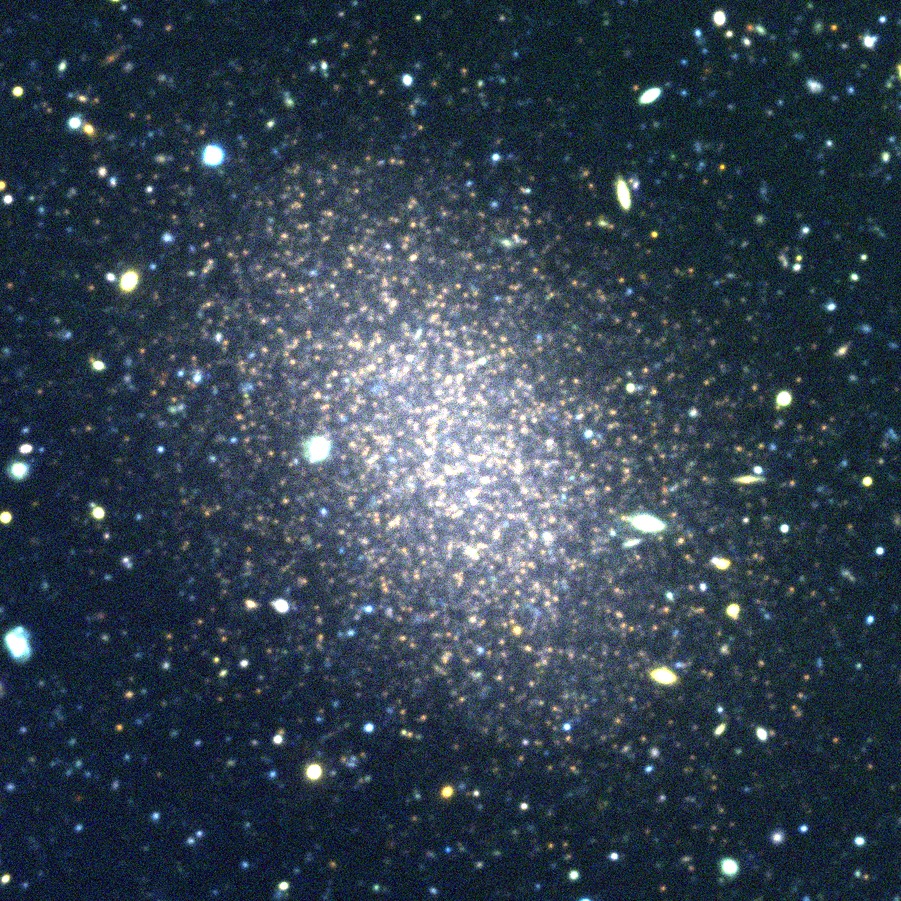}
  \includegraphics[width=.3\linewidth]{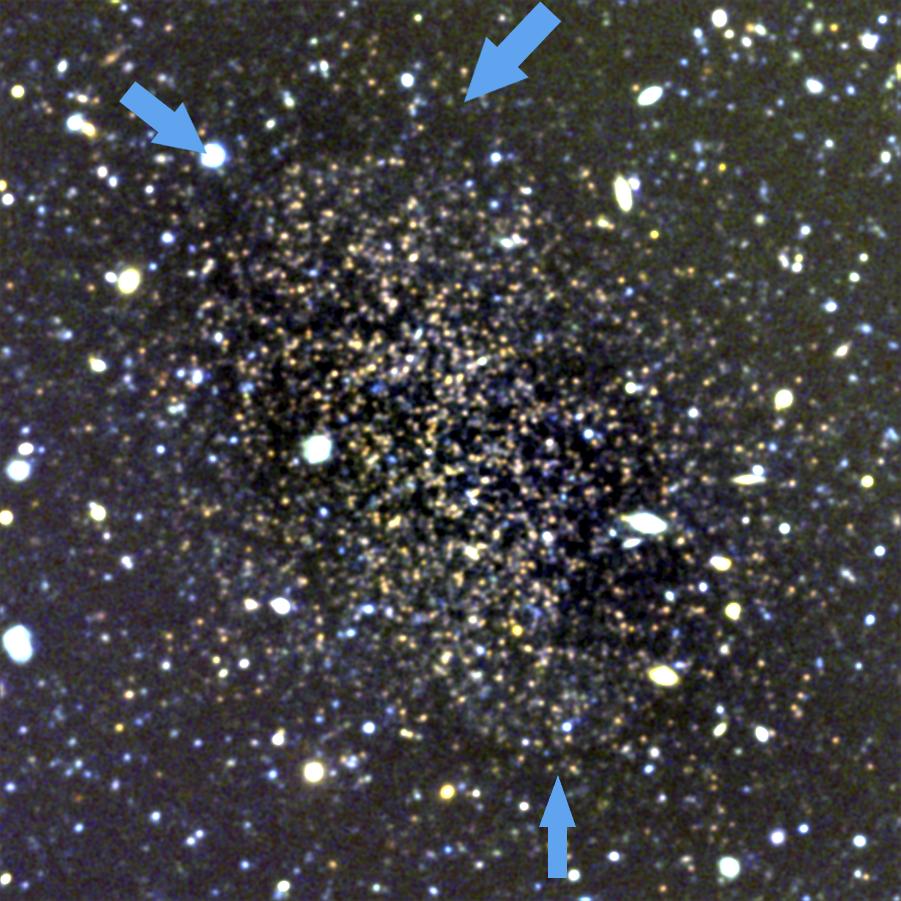}
  \caption{\emph{Left:} Example of the stellar distribution for a distorted S\`ersic profile, with a crossover radius of $80^{\prime\prime}$ and a position angle twist of $15\degree$. \emph{Center:} A composite RGB (\emph{R}, \emph{V}, \emph{B} bands) color image of a DDO 113 analog made with the profile in the left panel. \emph{Right:} The residuals after subtracting the best-fit symmetric \textsc{galfitm} model, smoothed with a $\sigma=1.5$ pixel Gaussian for better visibility. Arrows indicate the locations of model oversubtractions, resulting from the distortion.}
  \label{disrupted_sersic_galfit}
\end{figure*}

We generate one fake galaxy for each simple stellar population on each of three chips, avoiding the central chip due to the large angular size of NGC 4214. We find that the balance between resolved stars (mostly from the RGB) and diffuse light is most similar to DDO 113 for the 5 Gyr, $\text{[Fe/H]}=-1$ analog, although it is redder than DDO 113, while the best color match is the 1 Gyr, $\text{[Fe/H]}=0$ analog, though it is deficient in resolved RGB stars relative to DDO 113. Using HST data, \cite{Weisz2011} modelled the resolved CMD of DDO 113 and found that it had roughly constant star formation until about 1 Gyr ago when it quenched, and that it had formed $\sim50\%$ of its stars by 5 Gyr ago. The \textsc{galfitm} results for a fake 5 Gyr, $\text{[Fe/H]}=-1$ galaxy are shown in Figure \ref{galfitmodels_fake5gyr}.\par

\begin{figure}
\centering

  \includegraphics[width=\linewidth,page=1]{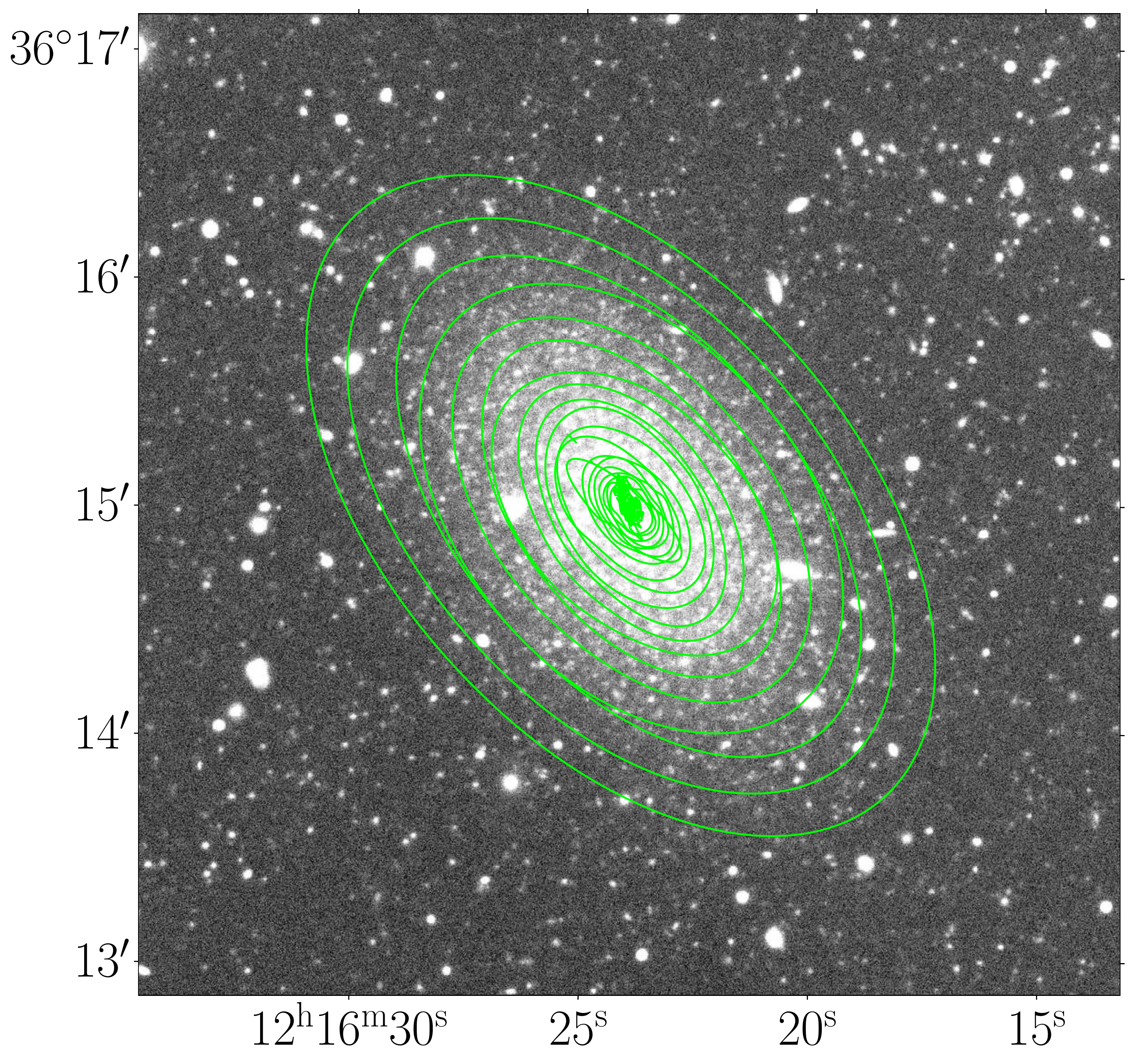} \hspace{0.06\textwidth}
  \includegraphics[width=\linewidth,page=2]{isophotes_py3_chip3_galfitm}
  \caption{\emph{Top:} The \emph{V} band elliptical isophotes for a DDO 113 analog. Distortions at small radii are due to the high density of resolved stars. \emph{Bottom:} \emph{V} band surface brightness profile of a DDO 113 analog, with the \textsc{galfitm} model and the intrinsic profile superposed. Unweighted and inverse-variance weighted S\`ersic fits to the isophote values are overplotted.}
\label{isophotes_plot}
\end{figure}

\begin{figure}
\centering
  \includegraphics[width=\linewidth,page=1]{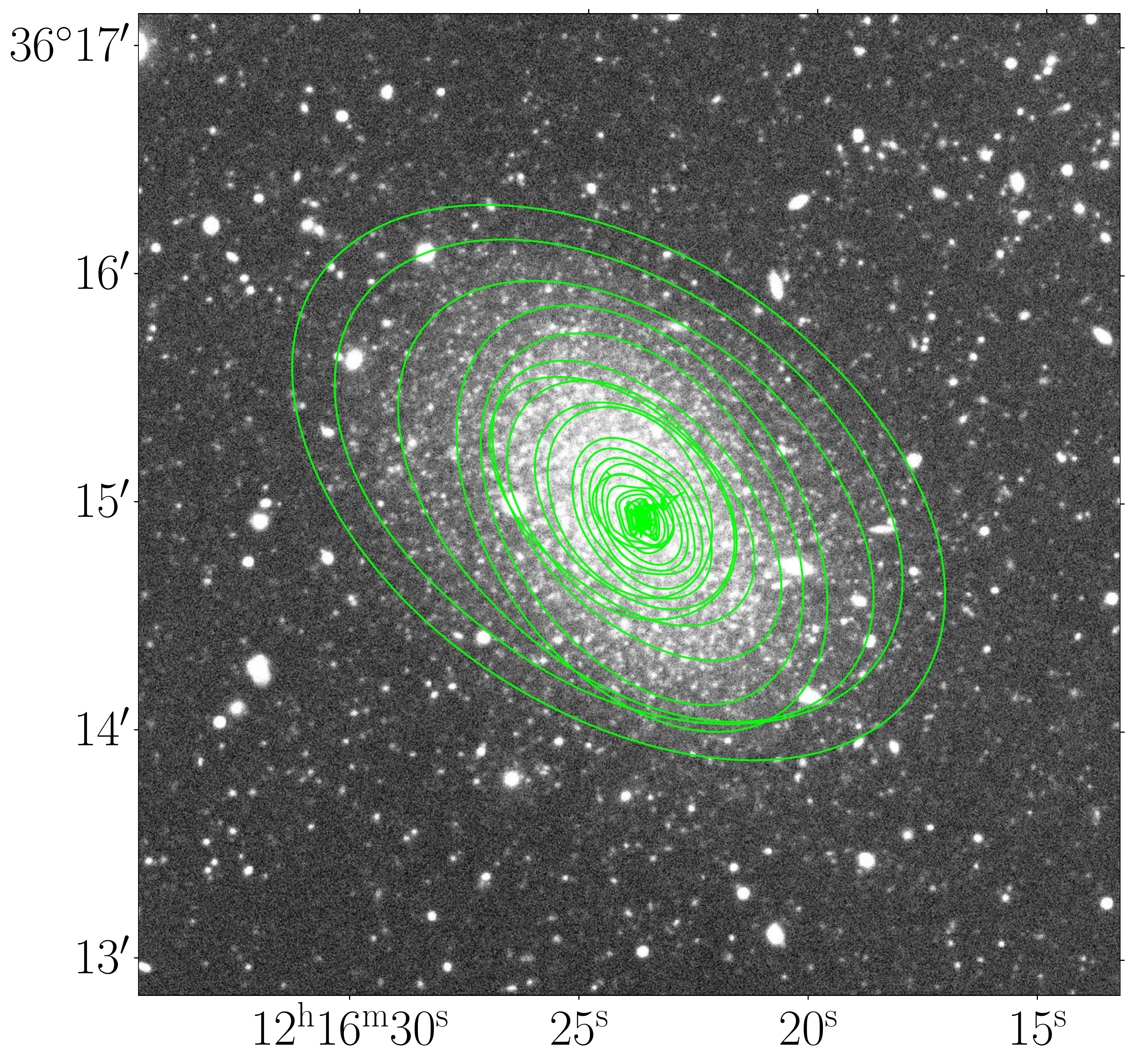}
  \caption{\emph{V} band elliptical isophotes for a simulated DDO 113 analog with a $15\degree$ position angle twist at a semi-major axis length of $80^{\prime\prime}$. }
  \label{disrupted_isophotes_plot}
\end{figure}

When we fit the fake galaxies with \textsc{galfitm}, the positions are correct to within $2^{\prime\prime}$, apparent magnitudes are correct to within $0.10$ magnitude, the effective radii are correct to within $1\farcs5$, the S\`ersic indices are correct to within $0.02$, the axis ratios are correct to within $0.05$, and the position angles are correct to within $2\degree$. The only notable differences between the results for the different isochrones were that the \emph{U} band magnitude and effective radius were poorly estimated for the 10 Gyr, [Fe/H]$=-2$ isochrone, likely due to its low luminosity and surface brightness. Similar to the real DDO 113, we observe no significant residuals after subtracting the \textsc{galfitm} models. \par

Next, we test our ability to detect tidal disruption by creating DDO 113 analogs with distorted S\`ersic profiles. The distorted S\`ersic profiles consist of two nested S\`ersic profiles, with identical morphological parameters except for the position angles, as might occur if the outer stellar distribution was subject to a tidal torque. Inside an ellipsoidal radius
\begin{equation}
  r_c=\sqrt{ (x \, \text{cos} \, \theta+y \, \text{sin} \, \theta)^2 +  \left( \frac{y \, \text{cos} \, \theta-x \, \text{sin} \, \theta }{q}\right)^2  }
\end{equation}

\noindent with an axis ratio $q$ and a position angle $\theta$ we sample from a S\`ersic profile with the structural parameters measured by \textsc{galfitm} and outside this crossover radius we sample from an identical profile rotated by $\Delta \theta$. We produce one fake galaxy on each of three chips using $r_c=80^{\prime\prime}$, roughly twice $r_e$ where $\mu_V\sim27$ mag $\text{arcsec}^{-2}$, and $\Delta \theta=15\degree$. An example of the stellar distribution, a color image of a fake galaxy, and the residuals from subtracting the best-fit axisymmetric \textsc{galfitm} model are displayed in Figure \ref{disrupted_sersic_galfit}.\par

With this method, finding distortions is easier with greater position angle twists and smaller crossover radii. Through testing models with different combinations of $r_c$ and $\Delta \theta$, we find that the combination of the smallest position angle twist and largest crossover radius that we can detect from the \textsc{galfitm} residuals is $\Delta \theta=10\degree$ with $r_c=100^{\prime\prime}$, where $\mu_V\sim28$ mag $\text{arcsec}^{-2}$.  \par

\subsection{Elliptical Isophotes} \label{Subsection:EllipticalIsophotes}

While using \textsc{galfitm} to search for tidal disruption in the body of the dwarf allows us to identify morphological asymmetries that could be indicative of tidal disruption, it is also desirable to look for changes in morphology as a function of distance from the dwarf's center quantitatively. For this we fit elliptical isophotes using the implementation from photutils v0.5 \citep{Bradley2018} of the method described in \cite{Jedrzejewski1987}. This method assumes the isophotes can be modelled as ellipses with harmonic distortions.   \par

Due to the presence of resolved stars, intensity fluctuations along the fitted isophotes are significant and can cause the fitting routine to return unreliable results. We perform sigma-clipping on the pixels used in each fitting iteration to mitigate this effect. At the beginning of each fitting iteration, the isophote's mean intensity and RMS error are measured and pixel values that are over $4\sigma$ from the mean intensity are ignored during the fit. We perform this operation twice for each fitting iteration. This clipping scheme does not completely remove the effects of resolved stars at small radii, where the point source density is high, but it improves fitting reliability and has minimal impact on isophotes at large radii and low surface brightness. We fit the elliptical isophotes in each band separately, initializing the iterative fitting process using the structural parameters from \textsc{galfitm}. We fix a minimum semi-major axis length $a=1^{\prime\prime}$ and set the isophotes to increase in size geometrically, by 15\% at each step. All isophotes have their best-fit harmonic distortions applied. Uncertainties on the surface brightness of each isophote are computed as the root-mean-square error of the pixel intensities. \par

\setcounter{figure}{7}
\begin{figure*}
  \centering
  \subfloat[\emph{U} band isophote fitting results.]{
  \hspace{0.03\linewidth}
  \includegraphics[width=0.4\linewidth,page=1]{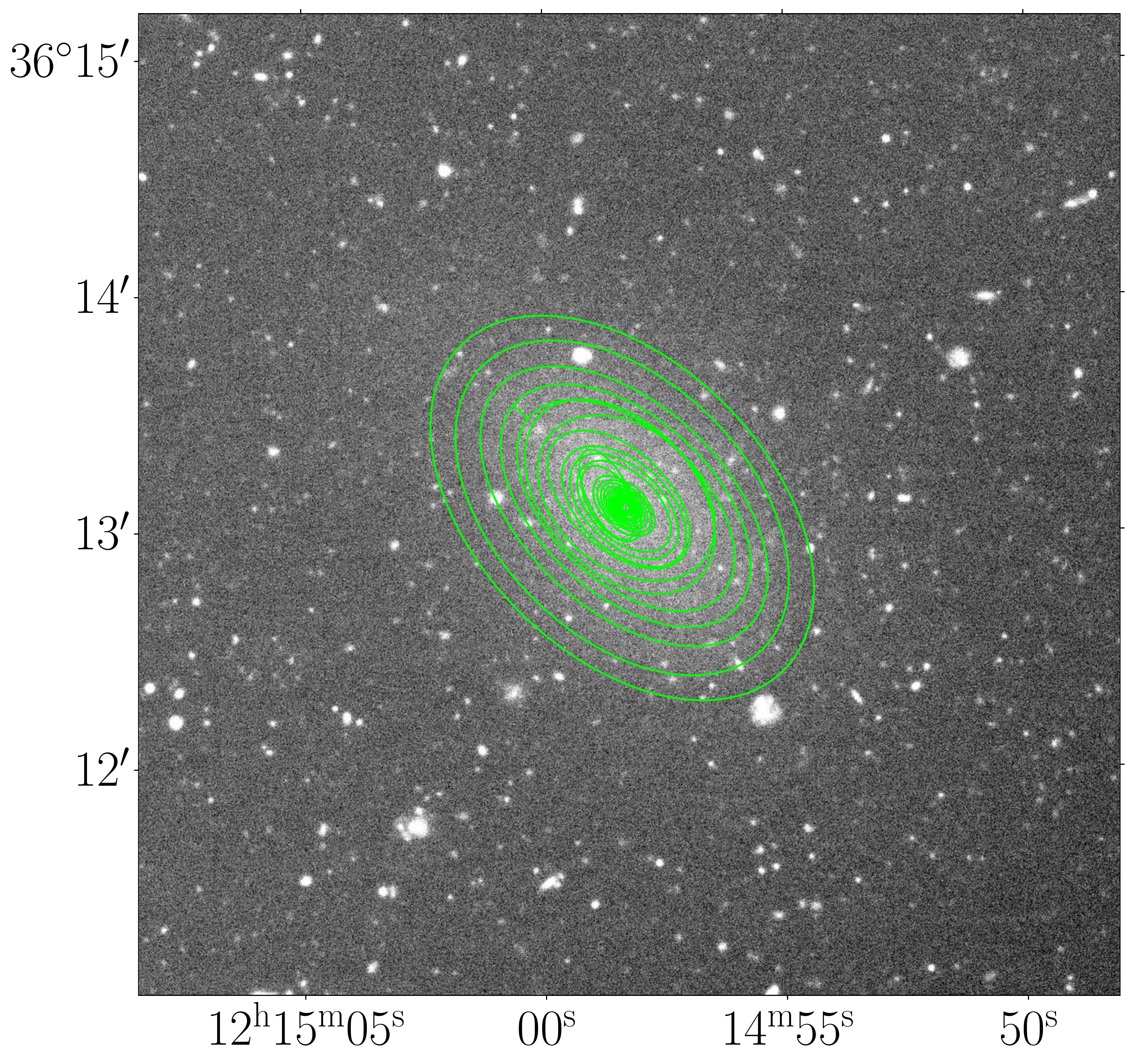} \hspace{0.1\linewidth}
  \includegraphics[width=0.4\linewidth,page=2]{isophotes_py3_ddo113} \hspace{0.15\linewidth}
  }
\newline
  \subfloat[\emph{B} band isophote fitting results.]{
  \hspace{0.03\linewidth}
  \includegraphics[width=0.4\linewidth,page=3]{isophotes_py3_ddo113} \hspace{0.1\linewidth}
  \includegraphics[width=0.4\linewidth,page=4]{isophotes_py3_ddo113} \hspace{0.15\linewidth}
  }
\newline
  \subfloat[t][\emph{V} band isophote fitting results.]{
  \hspace{0.03\linewidth}
  \includegraphics[width=0.4\linewidth,page=5]{isophotes_py3_ddo113} \hspace{0.1\linewidth}
  \includegraphics[width=0.4\linewidth,page=6]{isophotes_py3_ddo113} \hspace{0.15\linewidth}
  }
\end{figure*}
\begin{figure*}
\ContinuedFloat
  \centering
  \subfloat[\emph{R} band isophote fitting results.]{
  \hspace{0.03\linewidth}
  \includegraphics[width=0.4\linewidth,page=7]{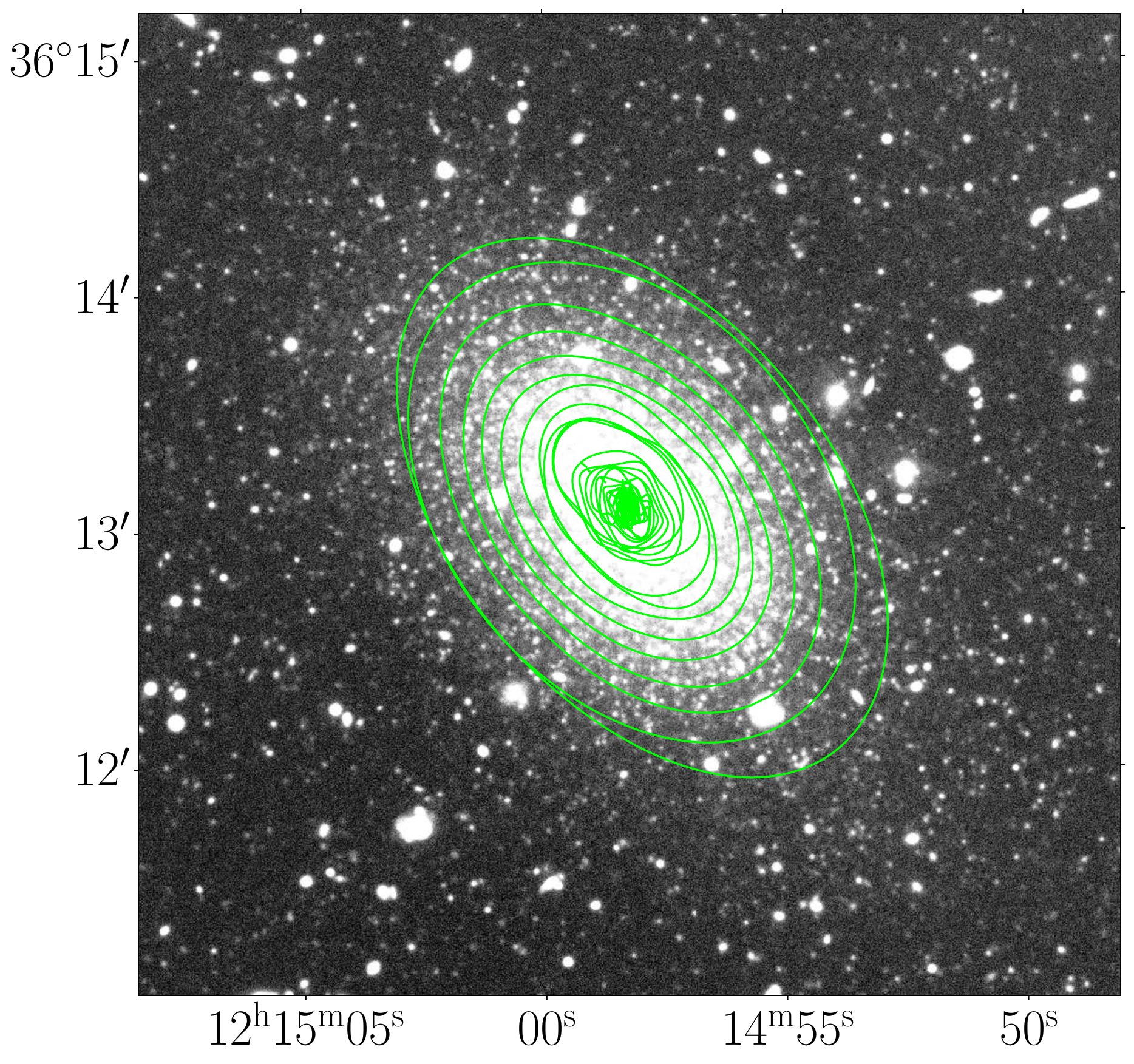} \hspace{0.1\linewidth}
  \includegraphics[width=0.4\linewidth,page=8]{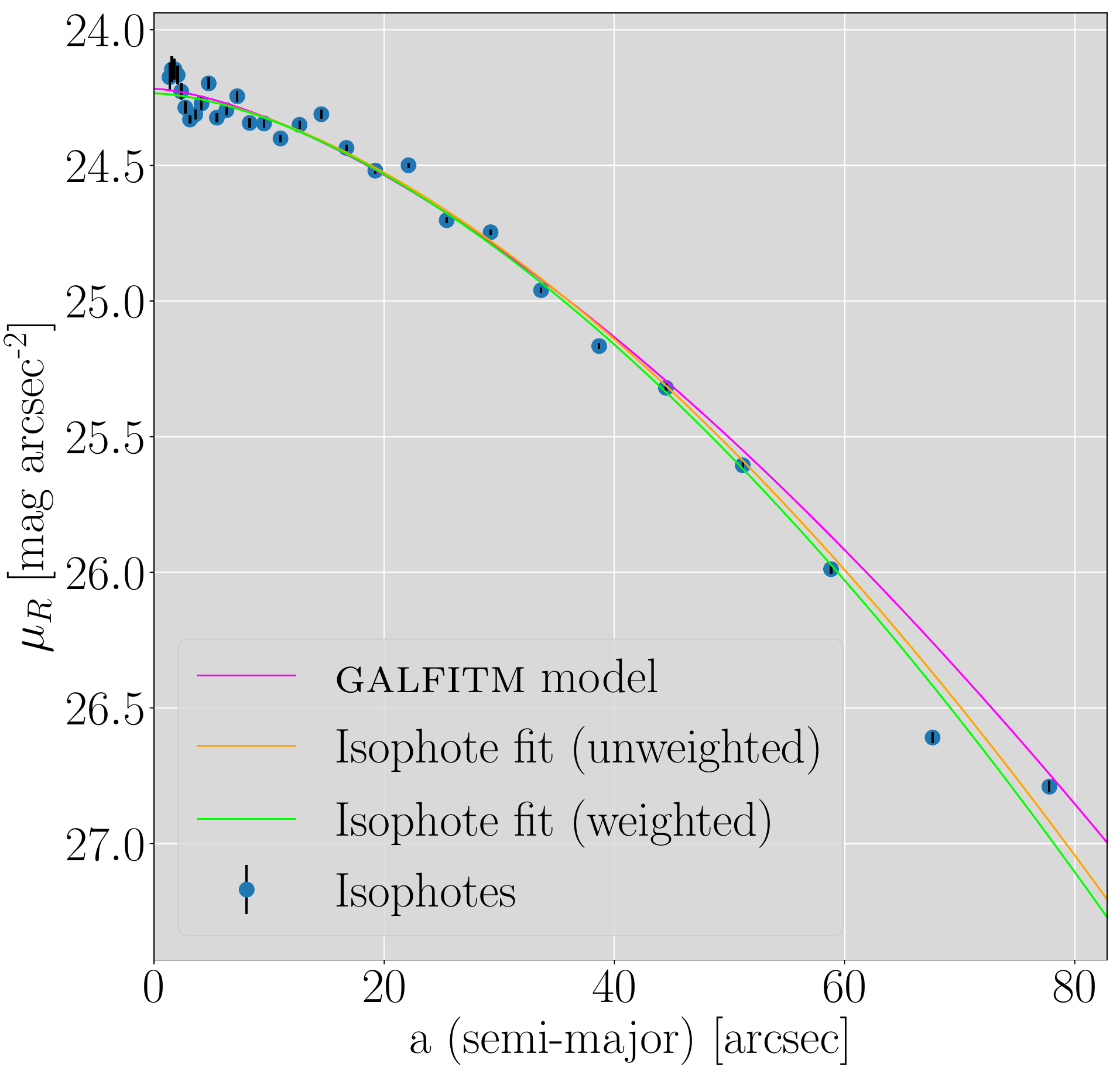} \hspace{0.15\linewidth}
  }
  \caption{\emph{From top left to bottom right:} Elliptical isophotes and corresponding surface brightness profiles of DDO 113 in \emph{U}, \emph{B}, \emph{V}, and \emph{R} bands. Elliptical isophotes have harmonic distortions applied and surface brightness profiles have best-fit \textsc{galfitm} models and both unweighted and inverse-variance weighted S\`ersic fits to the isophote values overplotted.}
  \label{figure:ddo113_isophotes_main}
\end{figure*}

We first examine the elliptical isophote results for the DDO 113 analogs discussed in \S \ref{subsection:galfit}. Over our full grid of these analogs, we find that the isophote fitting procedure produces accurate results with reasonable uncertainties for the $B$, $V$, and $R$ bands, while it has difficulty in $U$ band due to the lower surface brightness sensitivity of the data. Typical uncertainties are $5\degree$ for the position angle, 0.05 for the ellipticity, 0.10 for the magnitude, and $2\farcs2$ for the RA and Decl. An example of the fitted isophotes and surface brightness profile for an analog is shown in Figure \ref{isophotes_plot}. Most important for detecting tidal disruption are the position angle and ellipticity, with errors of $5\degree$ and $0.05$, respectively, in the outer regions, so we are sensitive to relatively small distortions. \par

We also fit isophotes to the distorted S\`ersic profile described in \S \ref{subsection:galfit} and shown in Figure \ref{disrupted_sersic_galfit} to assess our ability to detect tidal distortions. We use the same model setup as in \S \ref{subsection:galfit}, with a crossover radius of $80^{\prime\prime}$ and a position angle twist of $15\degree$. We detect this distortion easily, as shown in Figure \ref{disrupted_isophotes_plot}. We find that using elliptical isophotes allows us to detect slightly smaller distortion angles than using \textsc{galfitm}, but with reduced sensitivity at larger radii and lower surface brightness. The elliptical isophotes have the additional benefit of quantifying the distortion as a function of radius, while \textsc{galfitm} can only detect the presence of distortions. For the elliptical isophote method, we find the combination of smallest position angle twist and largest crossover radius that we are able to measure robustly is $\Delta \theta=7\degree$ at $r_c=90^{\prime\prime}$ where $\mu_V\sim27.5$ mag $\text{arcsec}^{-2}$.\par

With this testing concluded, we present the isophote fits and surface brightness profiles of DDO 113 in \emph{U}, \emph{B}, \emph{V}, and \emph{R} bands in Figure \ref{figure:ddo113_isophotes_main}. We note that the position angles of the outer isophotes in the fit for \emph{R} band were quite unstable to small perturbations in the parameters used to initialize the fit, such that the $2\degree$ uncertainty in the position angle from our \textsc{galfitm} fitting led to $\sim 15\degree$ changes in the isophote position angles at $a> 80^{\prime\prime}$. We therefore choose to reject these isophotes and keep only isophotes that are robust to changes in the initialization parameters within the \textsc{galfitm} uncertainties given in Table \ref{Table:galfitresults}. We hypothesize that the instability in \emph{R} band at large distances could be due to higher background point-source density compared to the other bands, which can affect the isophote fitting process even with the light sigma-clipping we apply to reject outlier pixels along each isophote. Uncertainties on the surface brightness profiles are root-mean-square errors of the intensities along the best-fit isophote and are likely underestimated due to the presence of resolved stars, background sources, and our two-iteration $4 \, \sigma$ clipping scheme. \par

The elliptical isophotes show no evidence of distortions to surface brightnesses of 26.5 mag $\text{arcsec}^{-2}$ in \emph{U}, 28.7 mag $\text{arcsec}^{-2}$ in \emph{B}, 27.2 mag $\text{arcsec}^{-2}$ in \emph{V}, and 26.7 mag $\text{arcsec}^{-2}$ in \emph{R}.\par

\subsection{Substructure Search with Resolved Stars} \label{subsection:resolved}
Although the integrated light profile of DDO 113 shows no signs of asymmetries that could indicate tidal disruption, it is possible that stars could be stripped from the exterior of the dwarf without significantly altering DDO 113's morphology in unresolved light. To investigate this, we search for substructures of resolved stars (e.g., streams) around DDO 113. We use the \textsc{allframe} point-source catalogs described in \S \ref{subsection:photometry}. \par

\begin{figure*}
  \centering
  \includegraphics[width=.42\linewidth,page=1]{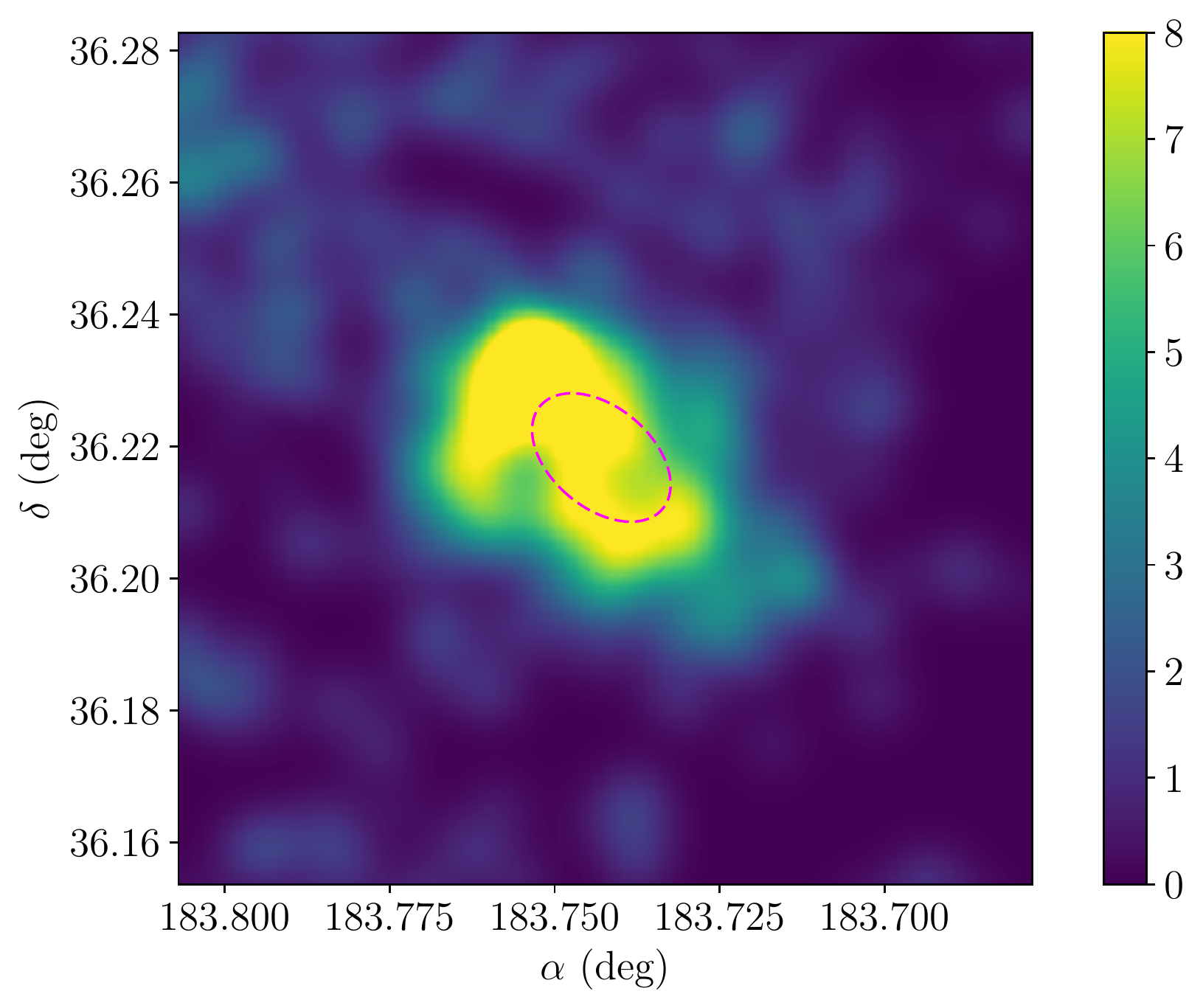} \hspace{0.05\textwidth}
  \includegraphics[width=.42\linewidth,page=1]{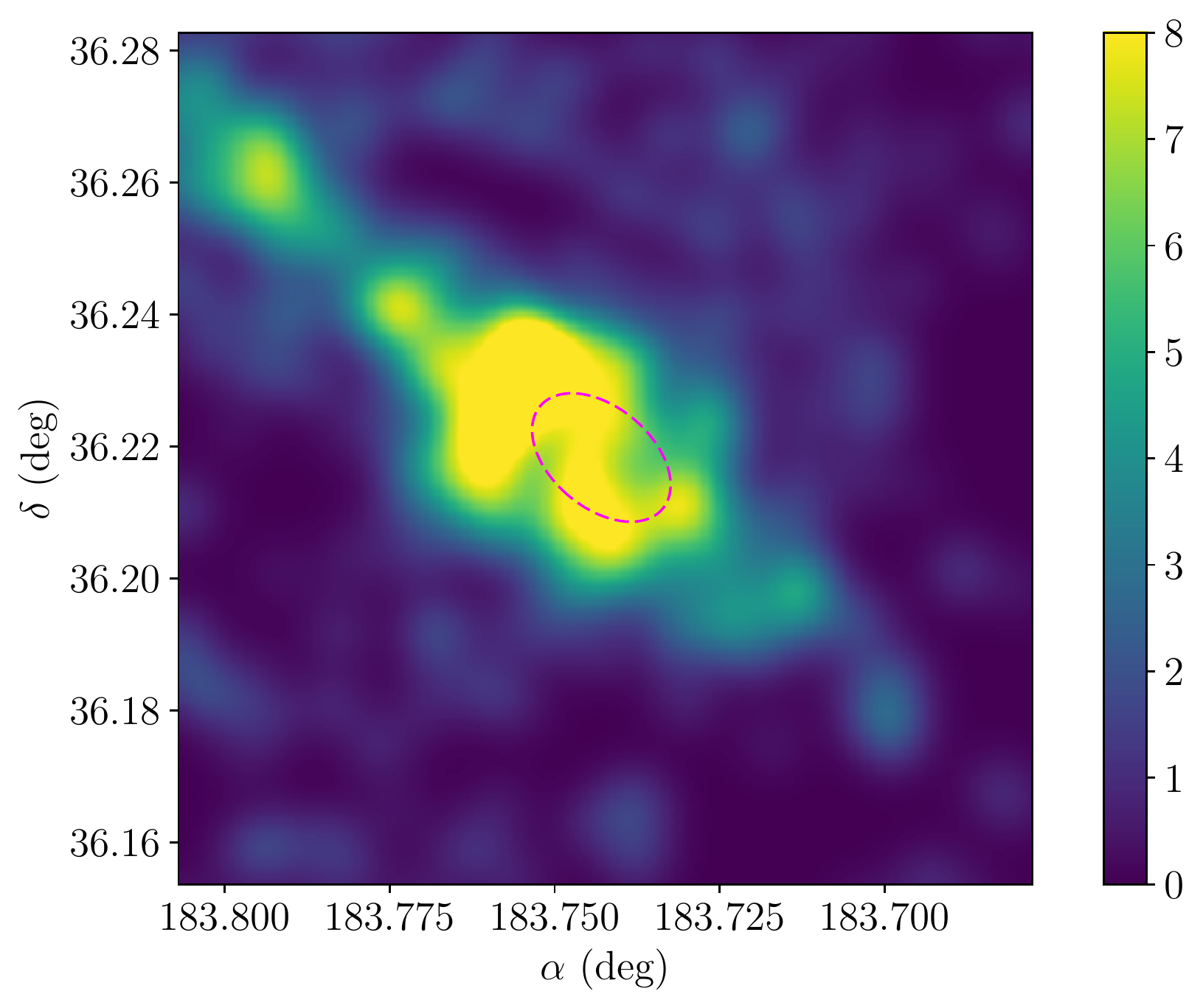}
  \caption{\emph{Left:} KDE map of the area around DDO 113 in resolved stars matching an old, metal-poor isochrone, smoothed on a $20^{\prime\prime}$ ($\sim 300$ pc) scale. An ellipse indicating DDO 113's effective radius is overplotted; we are largely incomplete inside this region. \emph{Right:} Same as left, but with a $\mu_V=29$ mag $\text{arcsec}^{-2}$ stream model added along the dwarf's major axis. This stream is easily detected, although its morphology in the map is irregular due to our reliance on stochastically distributed RGB stars. We can fully resolve the morphology for stream models with $\mu_V=28$ mag $\text{arcsec}^{-2}$.}
  \label{figure:cmd_map}
\end{figure*}

To reduce contamination from foreground MW stars and background galaxies, we filter our point-source catalogs with a 10 Gyr, $\text{[Fe/H]}=-2$ isochrone from PARSEC \citep{Aringer2009,Bressan2012,Chen2014,Marigo2017}. We considered other isochrone choices, but found that using an old isochrone maximized the signal from DDO 113 due to our ability to resolve the upper RGB. We used a two-color, likelihood-based weighting scheme, where the weight $w$ of the star $i$ is

\begin{equation}
  \begin{aligned}
  w_i= \, & \text{exp} \left[ - \ \frac{\left[ \ \left(B-R\right)_i - \left(B-R\right)_{iso} \ \right]^2}{2 \ \sigma_{(B-R)_i}^2} \right] \times \\
  & \text{exp} \left[ - \ \frac{\left[ \ \left(B-V\right)_i - \left(B-V\right)_{iso} \ \right]^2}{2 \ \sigma_{(B-V)_i}^2} \right],
  \end{aligned}
  \label{equation:weight}
\end{equation}

\noindent where $(B-R)_{iso}$ and $(B-V)_{iso}$ are the interpolated colors of the isochrone at a magnitude $R_i$, and the $\sigma$ are the uncertainties on the colors of star $i$ from the \textsc{allframe} photometry. Weights range between 0 and 1.\par

We use the weights to perform a weighted Gaussian kernel density estimation (KDE),

\begin{equation} \label{equation:kde}
  \hat{f}_h(\vec{x}) = \sum_{i=1}^{N} \ w_i \ \text{exp} \left[ \frac{- \left( \vec{x}-\vec{x_i} \right)^2}{h^2}    \right]
\end{equation}

\noindent where $\vec{x}$ is the position where the estimate is evaluated, the $\vec{x_i}$ are positions of the stars included in the estimate, $N$ is the total number of stars included in the estimate, and $h$ is the bandwidth of the kernel. A prefactor of $1/(\pi h^2 \sum_{i=1}^{N} w_i)$ can be added to the right side of Equation \ref{equation:kde} to give the classical normalization for a KDE, but we find that interpretation of the resulting maps is simpler without this normalization. In Equation \ref{equation:kde}, each kernel has a maximum value equal to its weight, such that the map values for the KDE can be interpreted analogously to a two-dimensional histogram, where a value of $\hat{f}_h(\vec{x})=2$ can be produced by two coincident stars at position $\vec{x}$ with $w=1$, for example. We include only stars with $w>0.2$ to cull stars with low membership probabilities, and require $|\vec{x}-\vec{x_i}|<3h$ for computational efficiency. \comment{the latter truncation means that

\begin{equation}
  \frac{1}{\pi h^2} \int_{-3h}^{3h} \text{exp} \left[ \frac{- \left( \vec{x}-\vec{x_i} \right)^2}{h^2}    \right]  \ d\vec{x} = \text{erf}(3)^2 \approx 0.999956
\end{equation}

\noindent which is a minimal effect.}

We evaluate $\hat{f}_h(\vec{x})$ on an $800^2$ grid with a linear pixel size of $1\farcs725$ and use a kernel bandwidth of $h=20^{\prime\prime}$ (or $\sim 300$ pc in projected distance). \comment{$h=24.5^{\prime\prime}$ and $h=10.5^{\prime\prime}$ (or $\sim 350$ pc and $\sim 150$ pc in projected distance, assuming a distance to NGC 4214 of 2.95 Mpc from \citealt{Dalcanton2009}).} We present our results in the left panel Figure \ref{figure:cmd_map}. We observe a slight enhancement in the resolved stellar density just outside the effective radius to the northeast; this is moderately visible in the images (e.g., Figure \ref{galfitmodels}) as an excess of resolved RGB stars. We do not believe this is a tidal feature, as it is large in scale and lies at a fairly small radius where our \textsc{galfit} and elliptical isophote measurements show no significant deviations from a symmetric S\`ersic profile in the extended, unresolved emission from faint stars. \comment{This feature could instead be explained by \comment{completeness effects due to the high point-source density, or} stochastic RGB sampling, as we are working with small numbers, or by internal dynamics \citep[e.g.,][]{El-Badry2016} if it is a legitimate feature.} \par

The right panel of Figure \ref{figure:cmd_map} shows a reprocessing after injecting a stream model with a constant surface brightness of $\mu_V=29$ mag $\text{arcsec}^{-2}$ along the major axis of DDO 113. The stream is easily detected, although its morphology appears irregular due to the stochastic distribution of bright RGB stars. For a stream model with $\mu_V=28$ mag $\text{arcsec}^{-2}$, the stream looks much more regular. We conclude DDO 113 has no associated tidal streams of reasonable size down to at least $\mu_V=29$ mag $\text{arcsec}^{-2}$. \par

We also tested detection of smaller substructures with \comment{sizes between $50 \ \text{arcsec}^{2}$ and $80 \ \text{arcsec}^{2}$ (or $715 \ \text{pc}^2$ and $1144 \ \text{pc}^{2}$)} radial scale lengths between 25 and 40 arcsec (or 358 and 572 pc), more analogous to tidal debris, and find that we can reliably recover these features with $\mu_V=28$ mag $\text{arcsec}^{-2}$, while we recover $\sim50\%$ of features with $\mu_V=29$ mag $\text{arcsec}^{-2}$. All the injected substructures that we recover appear larger in scale and more regular than any of the features outside the body of DDO 113 in the left panel of Figure \ref{figure:cmd_map}, further proving that DDO 113's stellar population has not been tidally stripped. \par

\subsection{Summary of Tidal Feature Search}
We used \textsc{galfitm} \citep{Peng2010,Haeusler2013} and elliptical isophote fits \citep{Jedrzejewski1987,Bradley2018} to measure the morphology of DDO 113 as a function of radius, and tested our ability to detect tidal features by analyzing simulated DDO 113 images based on both smooth and distorted S\`ersic profiles. Our \textsc{galfitm} fitting results and associated uncertainties are given in Table \ref{Table:galfitresults}\comment{ and our elliptical isophote uncertainties are given in Table \ref{Table:isophote_errors}}. Isophote fits and the surface brightness profiles of DDO 113 are shown in \comment{Figure \ref{figure:b_band_isophotes} and the results for our other bands are presented in} Figure \ref{figure:ddo113_isophotes_main}. We find no significant deviations from a standard S\`ersic profile with either method. \par

While it is possible that DDO 113 could exhibit signs of disruption beyond our limiting surface brightnesses of 28.7 mag $\text{arcsec}^{-2}$ in \emph{B}, based on isophote fits, \comment{\footnote{Based on elliptical isophotes; see \S \ref{Subsection:EllipticalIsophotes} and Appendix \ref{appendix:isophotes}.}} and $\sim$ 28 mag $\text{arcsec}^{-2}$ in \emph{V} and \emph{R}, based on the \textsc{galfitm} models, this region extends to twice the effective radius of DDO 113 and contains $\sim 91\%$ of DDO 113's luminosity. We found no evidence for associated streams in resolved stars around DDO 113 down to a limiting surface brightness of 29 mag $\text{arcsec}^{-2}$ in \emph{V}. Our limiting surface brightness is comparable to or fainter than the surface brightness of the streams associated with ultra-diffuse dwarfs of similar absolute magnitude to DDO 113 recently discovered in \cite{Bennet2018}, so we would have discovered any analogous features associated with DDO 113. These factors \comment{, coupled with the non-detection of tidal debris in the \textsc{galfitm} residuals,} lead us to conclude that DDO 113 shows no photometric evidence of tidal disruption that could explain the quenching of its star formation.\par

\section{Infall Time} \label{section:illustris}
With no signs of tidal disruption in our deep photometric data, we must look to other mechanisms to explain the quenching of DDO 113's star formation. One possibility is that DDO 113 has been environmentally quenched by its host, NGC 4214.  This possibility is especially attractive because of previous studies that show only a vanishingly small percentage of similar-mass objects in the field are quenched \citep{Geha2012}. In order to ascertain the likelihood of environmental quenching, and to isolate the primary sources of environmental quenching, we require an estimate for the time that DDO 113 has spent in NGC 4214's halo because the infall time sets the clock for quenching mechanisms. We look to cosmological simulations to find the infall time distribution of analogs of the NGC 4214-DDO 113 system, with physical properties matched to the observed system's present-day properties.  \par

Our approach is motivated by past work on determining infall time probability distribution functions for Milky Way satellites.  With data from the high-resolution $\Lambda$CDM Via Lactea II simulation \citep{Diemand2007,Diemand2008}, \cite{Rocha2012} showed that satellite infall times correlate strongly with their orbital energies. While orbital energies require 3D velocity information, \cite{Rocha2012} also showed that infall time estimates can be made based on the radius and radial velocity, as satellites in different stages of accretion populate different regimes of this parameter space. This approach for obtaining infall time estimates from the statistical properties of halos in cosmological simulations is promising for satellites which do not have well-constrained proper motions. \par

\comment{For satellites of the Milky Way, radial velocities are easily acquired by measuring line-of-sight (LOS) velocities and correcting for the Sun's orbital motion around the Milky Way, and Earth's orbital motion around the Sun. For extragalactic satellites, calculating radial velocities still requires 3D velocity information for both the host and the satellite. Here we will show that infall time estimates can be made using this method with only LOS velocities of extragalactic systems by mock observing systems similar to NGC 4214 and DDO 113 in cosmological simulations.\par}

As we will discuss below and in \S \ref{section:quenching}, DDO 113 lacks a robust radial velocity measurement despite being in the footprint of several 21-cm \HI emission surveys. \comment{Existing interferometric data are unable to reach the surface brightness required to detect DDO 113 \citep[e.g.,][]{Walter2008,Ott2012,Hunter2012}, and single-dish data with large beams ($>4^{\prime}$) show spatial and kinematic confusion with the disk of the host, NGC 4214 \citep[e.g.,][]{Tifft1988,Peek2011,Peek2018}. } As such, we look to other observables that are correlated with infall time to constrain our estimate for DDO 113. Our two most powerful observational constraints are 1) that DDO 113's projected distance from its host is only $\sim 9$ kpc\comment{-- a satellite that has been recently accreted is unlikely to be observed at such a small projected distance}; and 2) that DDO 113 shows no photometric evidence for tidal disruption, despite being at such a small projected distance\comment{; this will disfavor early accretions that have lost significant mass since accretion}. We use these constraints to identify analog systems in a cosmological simulation in order to derive an estimated infall time for DDO 113. \par

We utilize the public Illustris cosmological simulations \citep{Vogelsberger2013,Genel2014,Vogelsberger2014a,Vogelsberger2014b,Nelson2015,Rodriguez-Gomez2015}. Illustris simulates a $\Lambda$CDM cosmology with parameters from WMAP9 \citep{Hinshaw2013}. Illustris includes hydrodynamics, with a fiducial physics model \comment{that includes star formation using an ISM density criterion, stellar feedback in the form of kinetic outflows, chemical enrichment from supernovae and asymptotic giant branch stars, black hole seeding, growth, and merging, primordial and metal-line cooling, and both quasar- and radio-mode feedback from active galactic nuclei; the fiducial physics model is} presented in \cite{Vogelsberger2013}. We use the flagship Illustris-1 run, which simulates a comoving box of volume 106.5 Mpc\textsuperscript{3} with $1820^3$ particles each for dark matter, gas, and tracers that are used to track the Lagrangian evolution of the gas \citep{Genel2013}. \comment{The dark matter particles have a mass of $6.26 \times 10^6 \ M_{\odot}$ while the gas particles have a mass of $1.26 \times 10^6 \ M_{\odot}$. } Dark-matter-only versions of the Illustris simulation suite are also available, and we repeat our analysis using Illustris-1-Dark\comment{in Appendix \ref{Appendix:illustris-1-dark}}.\par

We utilize the friends-of-friends group catalogs\footnote{The ``groups'' identified in the friends-of-friends catalogs roughly correspond to halos, so we will use the two interchangeably, while we will use ``subhalos" to refer to entries in the \textsc{subfind} catalogs.}, \textsc{subfind} subhalo catalogs, and the \textsc{sublink} merger trees \citep{Rodriguez-Gomez2015} to identify halos and subhalos and track their evolution through time. The friends-of-friends halo finder requires a minimum of 32 particles, corresponding to $2 \times 10^8 \ M_{\odot}$ if the particles are all dark matter, while \textsc{subfind} requires a minimum of 20 particles that are gravitationally bound, corresponding to a mass limit of $1.3 \times 10^8 \ M_{\odot}$. DDO 113 should have a halo mass $\sim 10^{10} \ M_{\odot}$ using the abundance matching scheme of \cite{Moster2013}, which corresponds to about 1600 dark matter particles, so any DDO 113 analogs in Illustris are well above the dark matter mass resolution limit. \par

Given that DDO 113's stellar mass of $1.81 \times 10^{7} \, M_\odot$ \citep{Weisz2011} corresponds to $\sim 15$ star particles, \comment{and that \cite{Vogelsberger2014a} describes galaxies with $\text{M}_*=10^9 \ \text{M}_{\odot}$ as ``relatively poorly resolved,''} we do not attempt to identify DDO 113 analogs in Illustris-1 by stellar mass. Analogs for the NGC 4214 and DDO 113 system can be identified in Illustris-1 more reliably by using total halo masses, but doing so requires a mapping between stellar mass and halo mass. We adopt the subhalo abundance matching (SHAM) method as implemented in \cite{Moster2013}. SHAM matches an observational stellar mass function \comment{or luminosity function} to a simulated halo catalog assuming that the stellar-to-halo-mass relation is monotonically increasing. The relation of \cite{Moster2013} also seems reasonable compared to recent hydrodynamic simulations \citep[e.g.,][]{Hopkins2014}. \comment{The result is a mapping between stellar mass and halo mass that is grounded on the observed stellar mass function and the halo statistics of the simulation. The implementation of \cite{Moster2013} combines stellar mass functions from multiple surveys with varying depths to generate stellar mass functions from $z=0-4$, which allows them to construct a multi-epoch abundance matching scheme that varies the stellar-to-halo-mass relation with redshift.}  \par

\begin{figure}
  \centering
  \includegraphics[width=\linewidth,page=1]{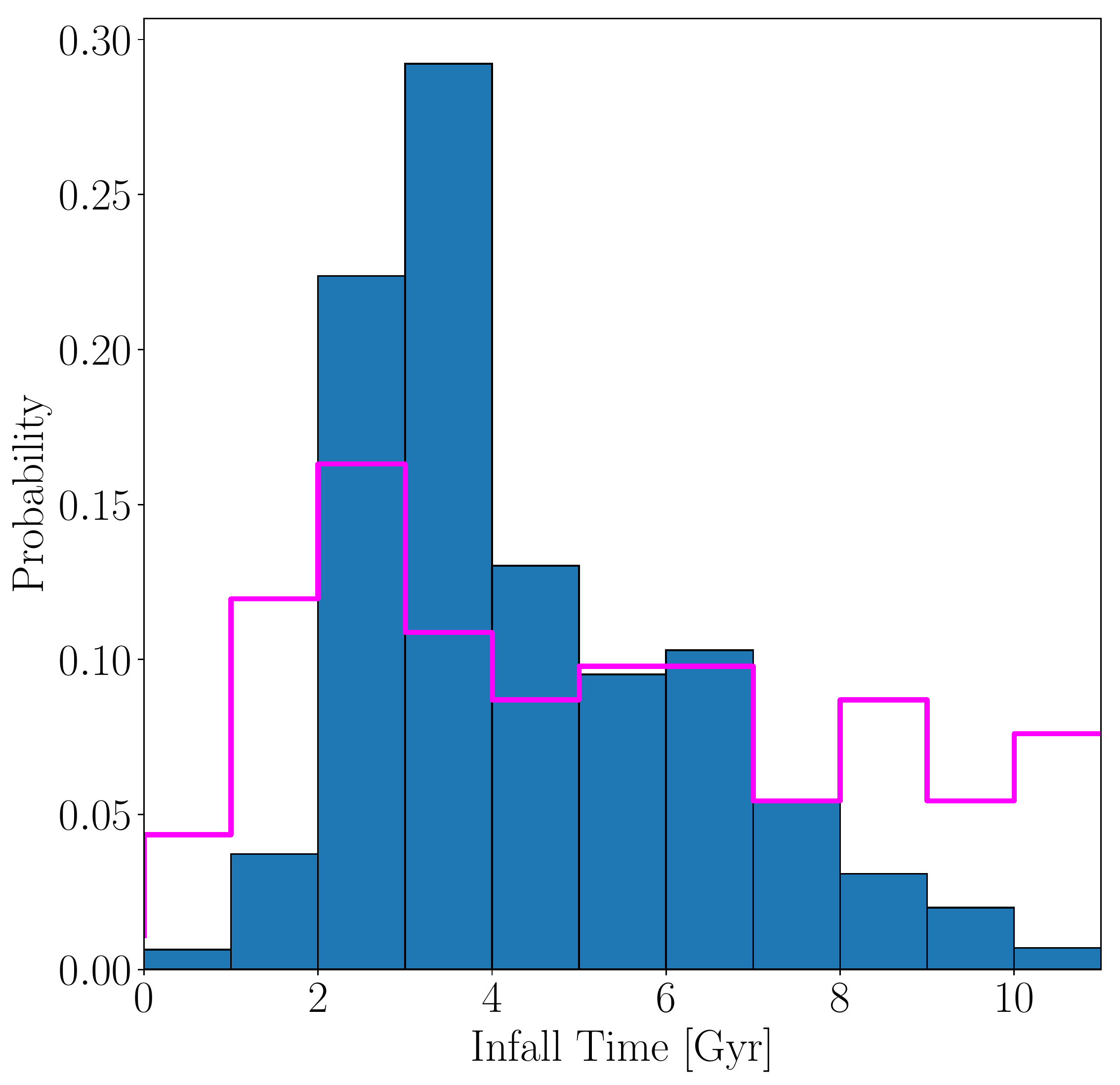}
  \caption{The blue filled (pink line) histogram shows the distribution of first infall times for 914 analogs of DDO 113 with (without) the tidal disruption criterion. Only systems that are randomly ``observed" at projected radii less than 20 kpc are included in the distribution. The distribution has a prominent peak between 2--4 Gyr, containing half the sampled systems, and a significant tail toward early infall times. Infall times from 0--2 Gyr and $>8$ Gyr are unlikely.}
  \label{figure:infalltime}
\end{figure}

Using this method, we look for groups where the most massive subhalo is close to NGC 4214's mass and the most massive satellite is close to DDO 113's mass. We use the total masses from the \textsc{subfind} catalogs for this purpose.\footnote{Often halo masses are reported as $M_{200}$, corresponding to the total mass enclosed in a sphere with mean density 200 times the cosmological critical density. This quantity is only defined for ``groups'' (halos) in the Illustris catalogs. Using these could include both host and satellite masses together, which is undesirable for our purpose, so we choose to use the \textsc{subfind} masses.} We begin by parsing the group catalog at $z=0$ to identify halos where the most massive subhalo (host) has a mass in the range $11 \leq \text{log}(M_{\odot}) \leq 11.5$, a range of masses centered on the \cite{Moster2013} abundance matching mass estimate for NGC 4214. \par

\begin{figure*}
  \centering
  \includegraphics[width=0.45\linewidth,page=2]{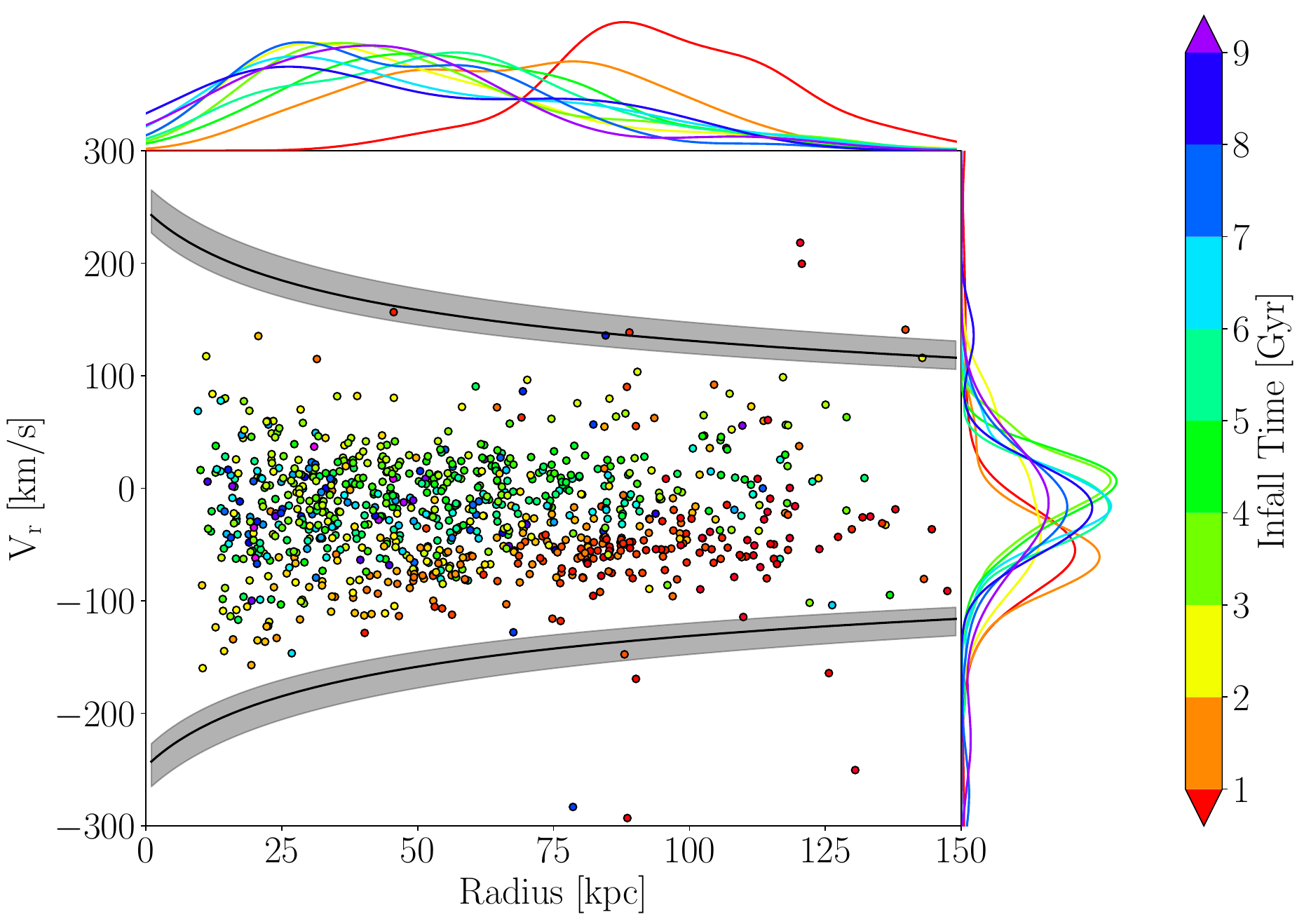} 
  \includegraphics[width=0.45\linewidth,page=3]{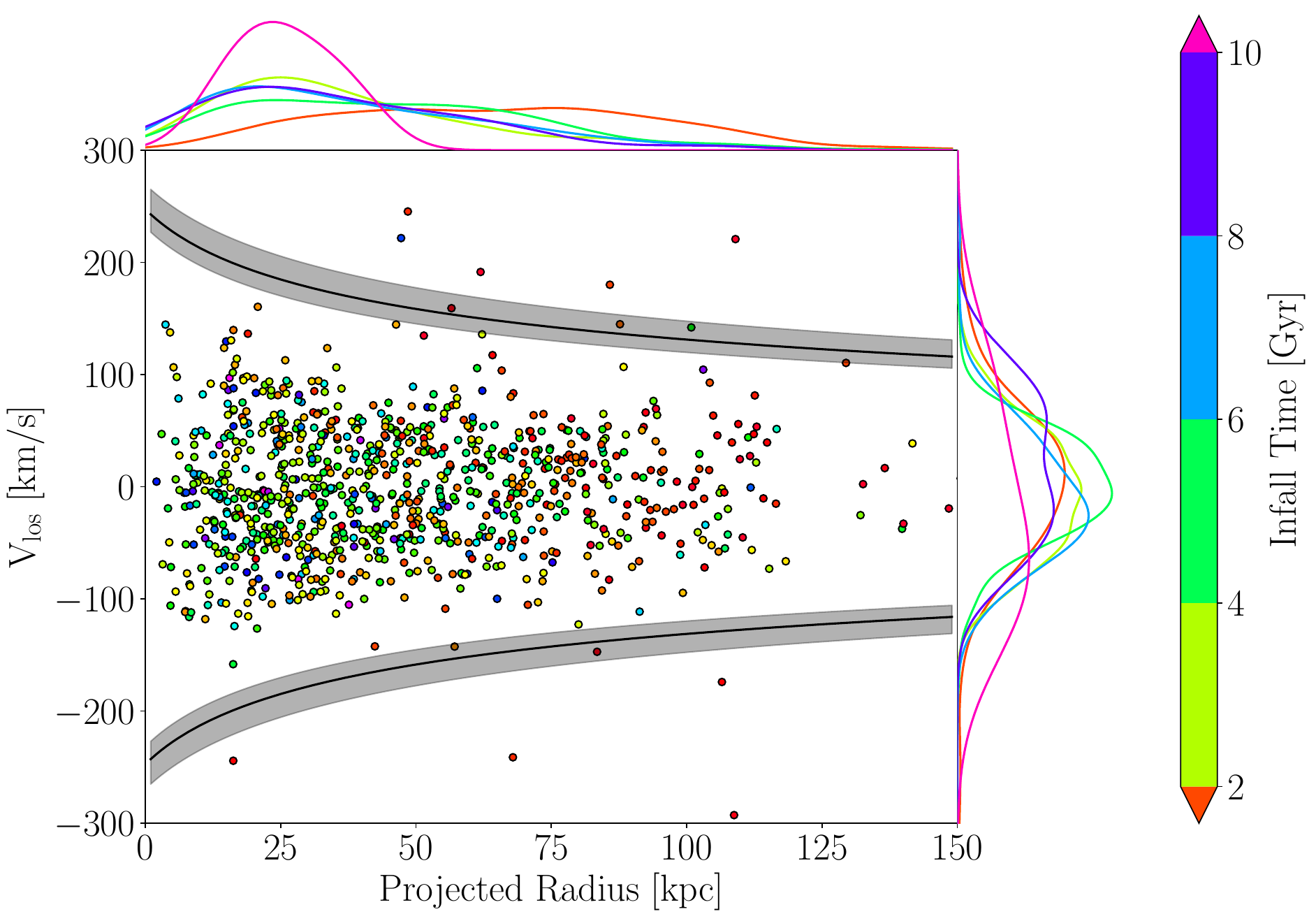} \\
  \includegraphics[width=.75\linewidth,page=4]{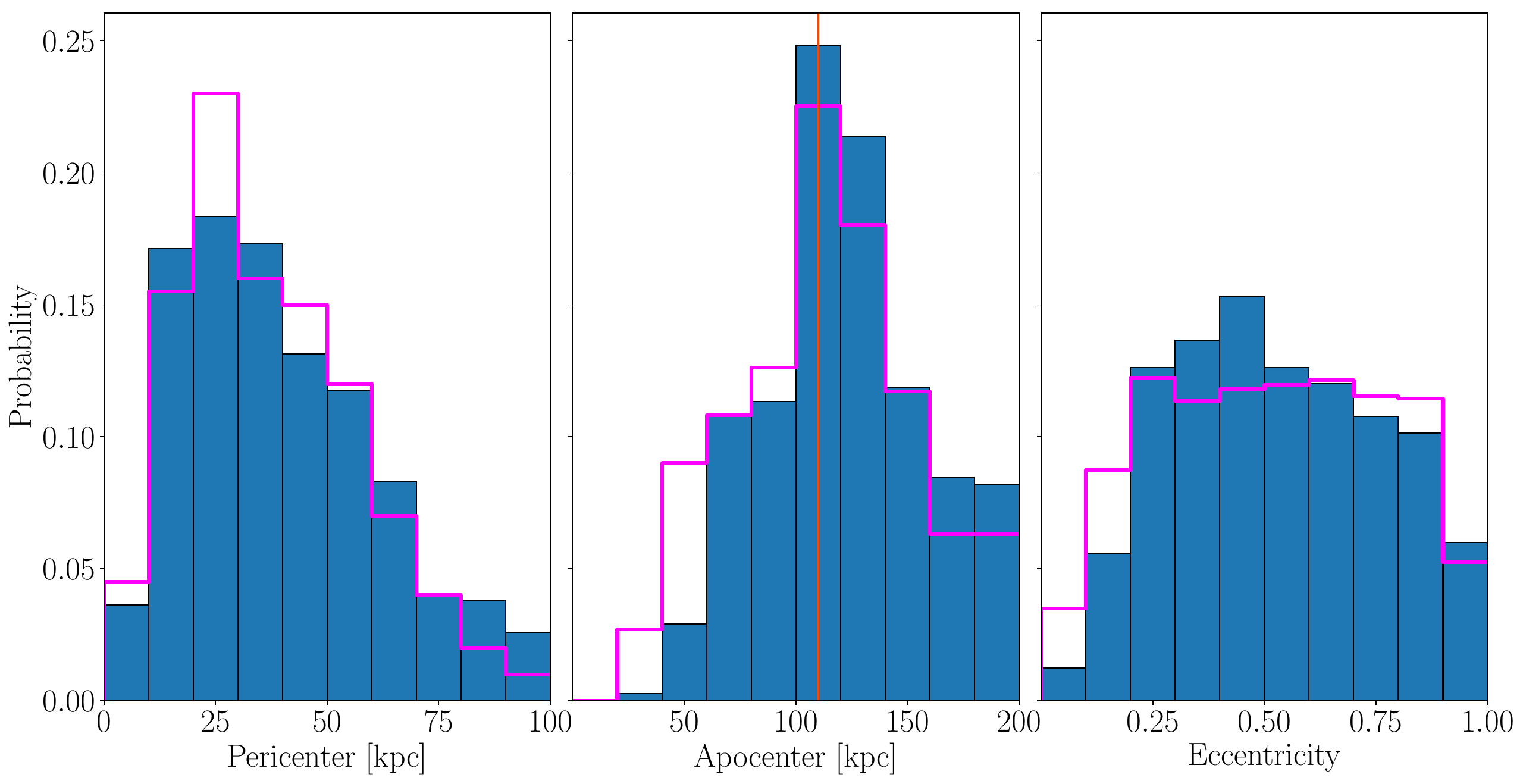}

  \caption{\emph{Top left:} Distribution of 914 DDO 113 analogs drawn from Illustris-1 in radial velocity and radius, with mean and $\pm 1\sigma$ escape velocity curves overplotted. \emph{Top right:} Distribution of the projected radii and line-of-sight velocities of the analog systems observed from a random direction. \emph{Bottom row:} Orbital parameter distributions for systems randomly observed with projected radii $<20$ kpc. Eccentricities are computed under the approximation of Keplerian orbits. Solid blue (pink line) histograms show distributions with (without) the tidal disruption criterion applied. The average virial radius of our analog hosts (110 kpc) is indicated by a vertical line on the apocenter plot.}
  \label{figure:illustris_orbital}
\end{figure*}

We track the subhalo populations associated with these groups through time using the \textsc{sublink} merger trees, and identify infall times based on the snapshot at which the subhalos transition from being part of a different group to belonging to the NGC 4214 analog groups. Often subhalos will have multiple infall events, where they are part of a group for a few snapshots, then become unassociated, then become associated again. For these subhalos, we define the infall time as the first snapshot at which the subhalo is associated with the group (first infall).\par

We identify DDO 113 analogs in these groups by requiring that the most massive satellites have a mass in the range $10 \leq \text{log}(M_{\text{infall}}) \leq 10.5$ at their infall time, corresponding to a range of masses centered on the \cite{Moster2013} abundance matching mass for DDO 113. We place this constraint on the infall mass rather than present-day mass as 1) the majority of a satellite's stellar mass is typically formed prior to accretion by its present day host, and 2) satellites are much more likely to experience tidal stripping than hosts, but the stellar component of a satellite is more centrally concentrated than its dark matter component, so the outer dark matter halo is stripped first \citep[e.g.,][]{Penarrubia2008}. This means that the stellar mass of the satellite is most closely correlated with its infall mass, which is typically also the maximum mass of the satellite over cosmic time. \cite{Penarrubia2008} used simulations to show that $\sim 90\%$ of a dark matter halo must be stripped before the stellar distribution is distorted, so we impose an additional criterion that the subhalos must have $M_{z=0} \ / \ M_{\text{infall}} \geq 0.1$ based on our photometric evidence that DDO 113 is not being tidally disrupted.\par

These criteria led to a sample of 914 systems. Next we randomly ``observed" these systems by projecting their positions along randomly-oriented observation vectors, keeping the infall times for cases with a projected separation $<20$ kpc, twice DDO 113's projected distance. Figure \ref{figure:infalltime} shows the distribution of infall times for 5000 random observations of systems selected both with and without the application of the tidal disruption criterion. From this distribution we conclude that DDO 113's first infall was likely 2--6 Gyr ago. \par

Application of the tidal disruption criterion significantly reduces the range of allowed infall times. In order to assess the origin of this improvement, we need to understand the orbital parameters of the satellites that we selected and discarded. 
With full 3D positions and velocities for our sample systems, we can estimate the satellite orbital pericenters and apocenters. We model the potentials of the hosts as spherically-symmetric NFW \citep{Navarro1996} profiles with concentrations derived from the power-law `stack, NFW' concentration-mass fit of \cite{Child2018}. For Illustris-1, we add an additional point-source potential with mass equal to the stellar mass of the host, although this only makes a difference for pericenters smaller than $\sim 20$ kpc. We also calculate orbital eccentricities under the approximation of Keplerian orbits. The orbital demographics of our analog systems are shown in Figure \ref{figure:illustris_orbital}. \par

Our mass-loss constraint $M_{z=0} \ / \ M_{\text{infall}} \geq 0.1$ removes many satellites at small apocenters that have experienced significant mass loss, typically over several orbits. Satellite apocenters are correlated with infall time, such that satellites accreted at earlier times have smaller apocenters due to the host having a shallower potential and smaller virial radius at earlier times. The removal of many satellites at small apocenters by the tidal disruption criterion is likely due to gradual mass removal from satellites accreted early. This is seen in the infall time distribution of Figure \ref{figure:infalltime}, where satellites with infall times $> 8$ Gyr are scarce after application of the tidal disruption criterion. Interestingly, the tidal disruption criterion appears to have little impact on the pericenter distribution. Satellites with smaller pericenters will encounter greater tidal forces, and should experience greater mass loss per orbit than equivalent satellites with larger pericenters. We observe $\sim 20\%$ fewer satellites with pericenters $< 10$ kpc after application of the tidal disruption criterion, but there are relatively few analogs with such small pericenters to begin with -- this may be because late-time infalls with such small pericenters are rare, or because satellites with radial infalls are quickly disrupted and do not survive to the present day. \par

\begin{figure*}
  \centering
  \includegraphics[width=.3\linewidth,page=1]{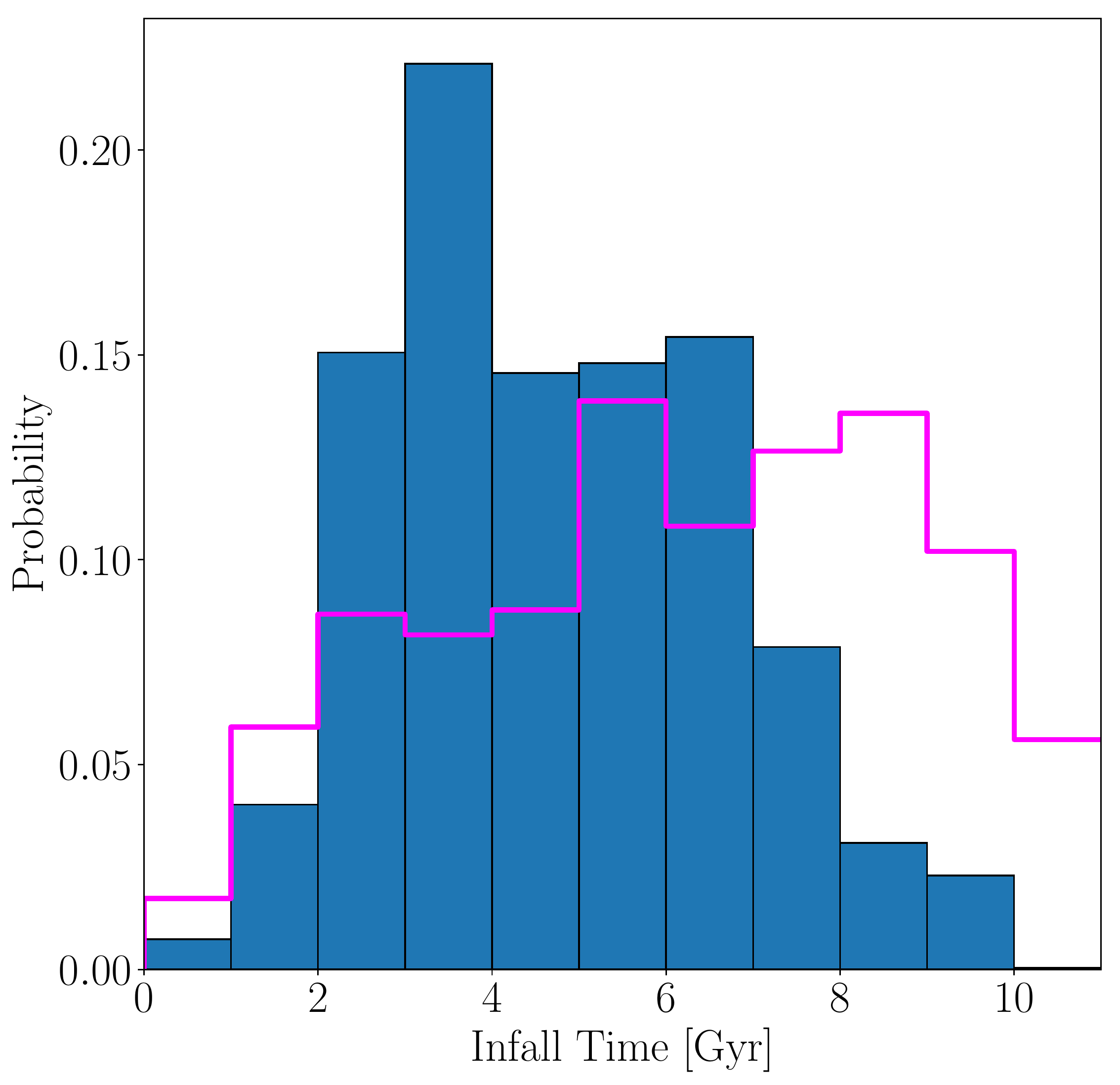}  \hspace{.05\textwidth}
  \includegraphics[width=.6\linewidth,page=2]{illustris_dark_analysis}
  \caption{\emph{Left:} As in Figure \ref{figure:infalltime}, but for analogs from Illustris-1-Dark. The distribution has more satellites accreted from 4--7 Gyr than found in Illustris-1, which may be explained by reduced tidal disruption in the dark-matter-only simulation. \emph{Right:} As in the bottom row of Figure \ref{figure:illustris_orbital}, but for Illustris-1-Dark. These mirror the Illustris-1 distributions closely, with the exception that satellites at small apocenters are not removed by the tidal disruption criterion as aggressively as we observe in Illustris-1.}
  \label{figure:illustris-dark}
\end{figure*}

A constraint on the line-of-sight velocities for the satellites can be added to this procedure easily and would improve our selection of analogs due to the correlation between orbital energy and infall time \citep{Rocha2012}. We originally intended to use the velocity of $284 \pm 6$ km $\text{s}^{-1}$ assigned to DDO 113 in NED\footnote{The NASA/IPAC Extragalactic Database (NED) is operated by the Jet Propulsion Laboratory, California Institute of Technology, under contract with the National Aeronautics and Space Administration.}, but the original measurement is based on GBT \HI~21-cm observations with a large ($\sim 9^{\prime}$) beam by \cite{Tifft1988}. Given the small projected distance of DDO 113 from NGC 4214 of $10\farcm6$, we were concerned that NGC 4214's gas disk could be spatially blended with DDO 113 and produce a false detection. This seems likely since the velocity assigned to DDO 113 is consistent with NGC 4214's velocity of $291 \pm 3$ km $\text{s}^{-1}$. We examined data from GALFA \citep{Peek2011,Peek2018}, which used the Arecibo Observatory with a $\sim 4^{\prime}$ beam, and found that the spectrum at DDO 113's position shows definite confusion in both position and velocity with the disk of NGC 4214. As such, we do not believe the GBT or GALFA data have sufficient spatial resolution to separate DDO 113's emission from the disk of NGC 4214. The high-resolution interferometric surveys VLA-ANGST \citep{Ott2012}, THINGS \citep{Walter2008}, and LITTLE THINGS \citep{Hunter2012} have data for DDO 113 with adequate spatial resolution to separate DDO 113's emission from the disk of NGC 4214, but all report non-detections. We obtained the THINGS and LITTLE THINGS data and attempted to extract spectra at the location of DDO 113 with no success. We estimate a similar upper limit of $4 \times 10^5 \, M_{\odot}$ to DDO 113's \HI~mass as that reported by VLA-ANGST \citep{Ott2012}. We adopt this upper \HI~mass limit for the remainder of the work. Since we have been unable to obtain a reliable velocity for DDO 113 we have not included any constraints on satellite kinematics when selecting analogs.\par

We repeat this analysis in full using the dark-matter-only Illustris-1-Dark run and find results that are qualitatively similar. The infall time and orbital parameter distributions are shown in Figure \ref{figure:illustris-dark}. The infall time is still most likely in the range of 2--6 Gyr, but there is an increase in early-accreted analogs relative to Illustris-1. We ascribe this to increased tidal disruption in Illustris-1 due to the presence of centrally-concentrated star particles in the host galaxy, which reduces the number of early-accreted satellites which survive to the present-day.\par 

From this analysis, we estimate a likely first infall time for DDO 113 of between 2--6 Gyr ago. While this is a wide range, it is significant that we have excluded a recent infall, as an infall time of less than 2 Gyr would require a very fast-acting quenching mechanism in order to suppress star formation 1 Gyr ago, as is observed \citep{Weisz2011}. DDO 113's small projected distance from its host and undisturbed stellar profile provide strong constraints on its infall time, such that we are now able to proceed with assessment of quenching mechanisms with a good estimate of the relevant timescale for quenching. This analysis appears robust to the treatment of the baryons, as the infall time constraints from the hydrodynamic Illustris-1 simulation are only moderately tighter than those obtained from the dark-matter-only sister simulation. It would be interesting to compare between simulations with different hydrodynamic treatments as well. \par

\section{Quenching Mechanisms}\label{section:quenching}
The goal of this section is to identify possible environmental quenching mechanism(s) for DDO~113 using the infall time estimate obtained in the previous section. We have a number of clues to help us solve this mystery. We begin by considering clues from DDO~113 itself. First, we have a high-fidelity SFH from the ANGST survey \citep{Weisz2011}, which shows not only when DDO~113 was quenched but also the star formation leading up to the end. The quenching is further confirmed by a lack of H$\alpha$ emission \citep{Kaisin2008}. Second, a viable quenching mechanism must be able to explain the lack of 21-cm \HI~emission \citep{Walter2008,Hunter2012,Ott2012} while maintaining the regularity of DDO 113's stellar distribution. In the field, galaxies with DDO~113's stellar mass have $\sim 1-6$ times more mass in \HI than in stars \citep{Papastergis2012,Popping2015,Sardone2019}, meaning that DDO~113 would have a gas mass of $M_{\text{H\textsc{i}}} \approx (2-12)\times 10^7 \, M_\odot$, well above the upper limit of $4 \times 10^5 \, M_{\odot}$ observed today \citep{Ott2012}. Third, we found no evidence of tidal stripping of stars (\S \ref{section:TidalDisruption}).  Finally, from our analysis of the Illustris simulations, we showed that it is very likely that DDO 113 fell into NGC 4214's halo 2--6 Gyr ago (\S \ref{section:illustris}). DDO 113 analogs in the field are predominantly star-forming \citep{Geha2012,Fillingham2018}, so this sets a minimum time over which environmental quenching mechanisms can act. \par

We also have clues from the host, NGC~4214, as to which quenching mechanisms may be at play.  While sub-L\textsubscript{*} galaxies are not expected to have accretion-shocked hot gas halos \citep{Birnboim2003,Keres2009}, there is strong observational and theoretical evidence that these galaxies generically have large, cool reservoirs of gas in their circumgalactic media \citep[CGM;][]{Bordoloi2014,Johnson2017,Fielding2017,Hafen2018}.  For galaxies like NGC 4214, the CGM is likely dominated by outflows rather than intergalactic medium inflow \citep{Hafen2018}.  Additionally, NGC~4214 is star-bursting, with noticeable holes in its \HI map \citep[e.g.,][]{Walter2008} and ionizing radiation leakage \citep{Choi2015}, which supports our hypothesis of an outflow-dominated CGM. Thus, environmental quenching processes (starvation and ram-pressure stripping) that traditionally have been ascribed to hot gas halos and ionizing radiation in higher-mass hosts (e.g., groups and clusters) may be at play with NGC~4214. \par

All of these lines of evidence additionally require that candidate quenching mechanisms act gradually, as DDO 113 only quenched within the last 1 Gyr, so it was likely still forming stars after first infall. This means candidate quenching mechanisms must act in the 1--5 Gyr between DDO 113's first infall and its quenching 1 Gyr ago. Removing the gas reservoir, ionizing the majority of the \HI, or increasing the scale length of the gas via heating (and thereby decreasing its mean column density) are all examples of mechanisms that could explain these observations without noticeably disturbing the stellar component of DDO 113. We use this timescale to investigate the possibilities of three different quenching mechanisms: strangulation, ram-pressure stripping, and tidal stripping. \par

\subsection{Strangulation}\label{subsection:strangulation}
Strangulation is the cessation of star formation due to the shut down of cold gas inflows when the dwarf moves from the field environment into the halo of a host \citep[e.g.,][]{Larson1980,Peng2015}. In this scenario, star formation should continue until the dwarf's \HI~reservoir is depleted by outflows and locked up in low mass stars. This mechanism depends on having a physical process by which a satellite galaxy can be disconnected from its circumgalactic gas reservoir (recycling) and intergalactic gas inflows (accretion). We ascribe the disconnection to NGC~4214's CGM, although we leave an investigation into the specific mechanism to future work. We roughly simulate the effects of strangulation using an instantaneous gas recycling and loss approach, which should be reasonable given the coarse time sampling of DDO~113's SFH (1 Gyr at best). We use the SFH of DDO 113 from \cite{Weisz2011} as input for our calculation. We calculate the \HI~mass loss rate as

\begin{equation} \label{equation: deltahi}
\Delta M_{\text{\HI}} = \eta \, \text{SFR} \, \Delta t + (1-R) \, \text{SFR} \, \Delta t
\end{equation}

\noindent where $\eta$ is the mass-loading factor and $R$ is the recycled fraction. In the first term, we assume all ejected gas is lost. In the second term, we account for both mass locked into stars and recycled from them back into the ISM through the recycled fraction $R$. We assume $R=0.3$ \citep{Portinari2004a}, although the exact value of $R$ is relatively unimportant for our calculation because of the large mass-loading factors. \par

We begin our calculation by assuming infall at a lookback time of 6 Gyr ($z=0.63$) when DDO 113's stellar mass is $10^7 \, M_{\odot}$ \citep{Weisz2011}, though we will later consider an infall 3 Gyr ago. We estimate DDO 113's \HI mass at infall with the power-law fit to $\langle M_{\text{gas}}/M_* \rangle$ given in \cite{Papastergis2012}, which implies an \HI mass upon infall of $5.6 \times 10^{7} M_{\odot}$ and that $M_{\text{\HI}}/M_* = 5.6$. 

We consider two relations for the mass-loading factors from \cite{Muratov2015} and \cite{Christensen2016}. The mass-loading factors of \cite{Muratov2015} are based on simulations using with the Feedback in Realistic Environments (FIRE) model \citep{Hopkins2014}, while the mass-loading factors of \cite{Christensen2016} are derived from simulations using \textsc{gasoline} \citep{Wadsley2004}. These relations bracket a reasonable range for the poorly constrained mass-loading factor \citep[e.g.,][]{Lu2015}. \par

\begin{table}
  \caption{The results of our strangulation calculation with the \citet[][M15]{Muratov2015} and \citet[][C16]{Christensen2016} mass-loading factors. SFR indicates the average star formation rate in $M_{\odot} \, \text{yr}^{-1}$ across the bins in lookback time. $\eta$ are the mass-loading factors calculated for each time bin with the two adopted relations. The $M\textsubscript{out} / M\textsubscript{\HI,0}$ rows indicate the fraction of initial \HI mass ejected over the time bins using the mass-loading factors in the row directly above.}
  \centering
    \hspace*{-4em}
    \begin{tabular}{c c c c c}
      \hline
      \hline
       & 3--6 Gyr & 2--3 Gyr & 1--2 Gyr & 0--1 Gyr\\
      \hline
      SFR & 0.00078 & 0.0025 & 0.0024 & 0.00036 \\
      $\eta$ (M15) & 21.3 & 17.2 & 15.5 & 14.1 \\
      M\textsubscript{out} / M\textsubscript{\HI,0} & 0.89 & 0.77 & 0.65 & 0.09\\
      $\eta$ (C16) & 3.5 & 3.5 & 3.5 & 3.5 \\
      M\textsubscript{out} / M\textsubscript{\HI,0} & 0.15 & 0.16 & 0.15 & 0.023\\
      \hline
    \end{tabular}
  \label{Table:strangulation_results}
\end{table}

Our results are summarized in Table \ref{Table:strangulation_results}. We find that using the mass-loading factors of \cite{Muratov2015} depletes the gas reservoir quickly ($\sim 2$ Gyr ago) while the mass-loading factor of \cite{Christensen2016} only depletes about half of the gas mass during the 6 Gyr since infall. Thus strangulation with the \cite{Muratov2015} mass-loading factors can explain the present day upper limit on DDO 113's \HI mass of $4 \times 10^5 \, M_{\odot}$ \citep{Ott2012}, while using the \cite{Christensen2016} mass-loading factors cannot. 

We can calculate the average mass-loading factor that depletes $\sim 90\%$ of the gas reservoir by a lookback time of 1 Gyr, when quenching began, by solving Equation \ref{equation: deltahi} for $\eta$ and averaging over the mass-loading factors and star formation rates from 1--6 Gyr,

\begin{equation}
\overline{\eta}=\frac{0.9 \, M_{\text{HI},0}}{\langle\text{SFR}\rangle \, \Delta t} + R -1
\end{equation}

\noindent where $\langle\text{SFR}\rangle$ is the average star formation rate over the elapsed time $\Delta t$. Using $M_{\text{\HI},0}=5.6 \times 10^{7} \, M_{\odot}$, $R=0.3$, $\Delta t=5$ Gyr, and $\langle \text{SFR} \rangle=0.00144 \, M_{\odot} \, \text{yr}^{-1}$, we find $\overline{\eta}=6.3$.  Repeating this analysis for an infall 3 Gyr ago yields $\overline{\eta}=11$. These values, calculated for first infall times that bracket our infall time distribution function, are between the values of \cite{Muratov2015} and \cite{Christensen2016} and in the range of $\eta=5-15$ favored by analysis of the CGM properties of sub-L\textsubscript{*} galaxies surveyed by COS-Dwarfs \citep{Bordoloi2014}. Thus, strangulation is a strong candidate for the dominant quenching mechanism of DDO 113. \par

This finding is unexpected given the relatively low mass of DDO~113's host, NGC~4214, with $M_\text{h}=1.7 \times 10^{11} \, M_{\odot}$ \citep{Weisz2011,Moster2013}. Strangulation is typically thought to remove the hot gas envelope of the satellite, preventing replenishment of the cold gas reservoir \citep[e.g.,][]{Larson1980}, but separation of the satellite from any significant gas inflows (e.g., pristine gas inflows from the cosmic web) is also necessary. Historically, strangulation has been considered only in the context of a host's hot gas halo. Recent numerical work indicates that halo masses $>2 \times 10^{11} \, M_{\odot}$ are required for formation of a hot, virialized gas halo \citep[e.g.,][]{Birnboim2003,Keres2009}, while analytic calculations yield a higher minimum halo mass of $> 10^{12} \, M_{\odot}$ \citep[e.g.,][]{Wang2008}. If strangulation is indeed the dominant driver of quenching for DDO~113, it is a first indication that the cool CGM may play a driving role in satellite quenching, as NGC~4214 lies below the hot gas halo formation regime.

\subsection{Ram Pressure Stripping}\label{subsection:rps}

Traditionally, ram-pressure stripping of gas from satellites has been understood in the context of hot gas halos in group- to cluster-scale environments \citep[e.g.,][]{Gunn1972,Tonnesen2009,fillingham2016}. In this section, we consider if a cool, outflow-driven CGM of NGC~4214 might remove DDO~113's \HI reservoir faster than by starvation. We calculate how much gas may be removed in a single orbit, and do not consider how the gas remaining in the satellite evolves, nor do we consider the interplay among starvation, outflows, and photoheating in combination with this effect.  Our goal is to determine how much gas may be stripped by ram-pressure stripping alone.

Gas may be removed from a satellite if the ram pressure from the host exceeds the gravitational restoring force per unit area of the satellite, 
\begin{eqnarray}
    \rho_\mathrm{h}(R) \, V_{\mathrm{sat}}^2(R) > \Sigma_{\mathrm{gas},s}(r) \, \frac{G M_\mathrm{s}(r)}{r^2}.
\end{eqnarray}  
Here, $R$ denotes the distance from the satellite to the host, and $r$ is the distance from a point in the satellite to the satellite center \citep{McCarthy2008,fillingham2016}.  The host galaxy's CGM has a density profile $\rho_\mathrm{h}(R)$, and the satellite's velocity with respect to the CGM is $V_{\mathrm{sat}}$.  This velocity may include a component due to the bulk motion of the CGM on account of its likely significant outflow component.  The satellite's cold gas surface density is $\Sigma_{\mathrm{gas},s}(r)$, and we treat the enclosed satellite mass $M_\mathrm{s}(r)$ as purely dark matter in origin, given the typically high mass-to-light ratios of dwarf galaxies \citep[e.g.,][]{Simon2007,Walker2007,Strigari2008a}.  Because the host's gas density peaks toward the center of the halo and the satellite's speed relative to the host is maximized at pericenter, ram-pressure stripping can, to first order, be considered to act dominantly at pericenter passage, and as an outside-in process.

We model the restoring force on DDO~113's gas in a manner analogous to the treatment of the Milky Way satellites in \citet{fillingham2016}. We use the infall gas mass estimate of $5.6 \times 10^{7} M_{\odot}$ derived in \S \ref{subsection:strangulation}. Following \citet{Popping2012,Popping2015}, based on work by \citet{Kravtsov2013}, we model the surface density of the cold gas as an exponential profile with a scale radius $\chi_{\mathrm{gas}} = 1.5-2.6$ times larger than the stellar half-light radius from \S \ref{subsection:photometry}.  We model the dark-matter halo of DDO~113 according to the SHAM relation of \cite{Moster2013} described in \S \ref{section:illustris}, with an average value of the halo concentration. 

We model the CGM of NGC 4214 with an isothermal profile, $\rho_{\mathrm{h}}(R) \propto R^{-2}$, as found in both staged and cosmological simulations \citep{Fielding2017,Hafen2018}. To normalize the density profile, we use the high-outflow case of \citet{Fielding2017}, which produces the highest CGM density of their models, $n_H = 10^{-3} \hbox{ cm}^{-3}$ at $R/R_{vir} = 0.1$. This is also consistent with the CGM density found in the cosmological simulations of \citet{Hafen2018}.  Lower mass-loading factors, consistent with the mass-loading factor inferred from a study of stellar and gas-phase metallicities \citep{Lu2015}, give a normalization a factor of 2 lower in the simulations of \citet{Fielding2017}.

For the satellite orbit, we consider the range $V_{\mathrm{sat}} = 150 - 300 \ \text{km s}^{-1}$.  The low value is close to the escape velocity at the high end of our pericenter distribution function (Fig.~\ref{figure:illustris_orbital}). The high value is greater than the escape velocity anywhere in the NGC 4214 halo, but allows for significant bulk motions in NGC 4214's CGM that are comparable to the escape velocity.  While this assumption is likely extreme, we consider it because NGC 4214 is star-bursting.

With these assumptions, we find it unlikely that ram-pressure stripping is the dominant source of quenching for DDO~113, although it may accelerate quenching.  The upper limit on the remaining cold gas mass of DDO~113 is at most only a few percent of the likely infall gas mass, if DDO~113 had field-galaxy properties at infall.  In order to strip $\gtrsim 95\%$ of the cold gas by ram-pressure stripping alone, we find that DDO~113's pericenter would have to be less than 10 kpc or the satellite velocity would have to be almost a factor of two higher than the escape velocity from NGC~4214, and the effective radius of DDO~113's cold gas would have to be high ($\chi_\mathrm{gas} \gtrsim 2.5$). Given the distributions of velocities and pericenters of the DDO 113 analogs in Illustris-1, neither the low pericenter nor the high satellite velocity are probable. For the peak of the satellite pericenter probability distribution shown in Figure \ref{figure:illustris_orbital} ($r_p = 25$ kpc), only about half the gas is stripped for a satellite traveling at the escape velocity through our highest-density choice for NGC~4214's CGM. Thus, ram-pressure stripping can be a contributor to gas loss, but is unlikely to be the main driver.

\subsection{Tidal Stripping}\label{subsection:tidal}
While the first two mechanisms depended on the presence of a host CGM, we can also estimate how much gas may have been stripped by gravitational means---tidal stripping.  Although the stellar component of DDO~113 appears undisturbed, cool gas typically extends beyond the optical radius \citep[e.g.,][]{Hunter2012}.  Thus, it is possible to tidally strip significant quantities of gas while leaving the stars largely untouched.  We use the definition of tidal radius $r_t$ employed by \citet{Penarrubia2008},
\begin{equation}
    \langle \rho_{sat}(r_t) \rangle = 3 \, \langle \rho_{host}(R) \rangle,
\end{equation}
where $\langle \rho_{sat}(r_t) \rangle $ is the dark-matter density within the tidal radius of the satellite, and $\langle \rho_{host}(R) \rangle$ is the average host dark-matter density within a radius $R$, which we take to be the pericenter of the satellite's orbit.  This definition of the tidal radius is equivalent to that derived for point masses in the circular restricted three-body problem.  \citet{Penarrubia2008} find that it takes multiple orbits for a satellite to lose the mass outside its pericenter-based tidal radius.  Given the short timescale allowed for quenching and gas removal, we likely overestimate the effect of tidal stripping if it were the only process acting on the gas.

With average choices for the halo masses and concentrations of the host and satellite, we find that the typical tidal radius at pericenter is $r_t \approx 5$ kpc, given the pericenter distribution in Figure \ref{figure:illustris_orbital}. If we model the cold gas disk of the satellite as in \S \ref{subsection:rps}, we find that only 5--20\% of the cold gas of DDO~113 can be tidally stripped if no other process acts on the gas.  Thus, tidal stripping alone does not explain the dearth of gas in DDO~113.

\section{Conclusion} \label{section:conclusion}
In this paper, we use deep LBT/LBC observations of DDO 113 in concert with simulations and analytic calculations to suggest for the first time that a low-mass galaxy's CGM quenched star formation in a Fornax-mass dwarf satellite galaxy.\par 

 We use \textsc{galfitm} and elliptical isophotes to search for deviations from regular S\`ersic profiles in our imaging data, with image simulations of both regular and tidally distorted DDO 113 analogs to quantify the systematic uncertainties in our measurements. We find that DDO 113's stellar distribution follows an undisturbed S\`ersic profile out to twice its half-light radius, with no detectable signs of irregularities in the light profile down to surface brightnesses of 28.7 mag $\text{arcsec}^{-2}$ in \emph{B} and $\sim$ 28 mag $\text{arcsec}^{-2}$ in \emph{V} and \emph{R}. We find no evidence for stellar streams associated with DDO 113 in resolved stars down to $\mu_V=29 \ \text{mag} \ \text{arcsec}^{-2}$. These results indicate that tidal disruption is likely not the cause of DDO 113's quenching. \par

To investigate other quenching mechanisms, we estimate an infall time for DDO 113 based on analog systems drawn from the Illustris-1 cosmological simulation \citep{Vogelsberger2013,Genel2014,Vogelsberger2014a,Vogelsberger2014b,Nelson2015,Rodriguez-Gomez2015}. We use halo mass constraints for both DDO 113 and its host, NGC 4214, to select analog systems, then simulate observations of the analog systems to leverage DDO~113's small projected distance ($\sim$ 9 kpc) from its host. We then apply a theoretically-motivated tidal disruption criterion \citep{Penarrubia2008} based on our observation that DDO 113's stellar component is undisturbed; this additional criterion significantly tightens our infall time estimate. Systems that meet the tidal disruption criterion and are mock observed at small projected distances are used to estimate DDO 113's infall time. From this analysis, we estimate that DDO 113's first infall was likely between 2--6 Gyr ago. \par

We use this timescale to investigate the plausibility of several environmental quenching mechanisms: tidal stripping, ram-pressure stripping, and strangulation. Intriguingly, we find that strangulation is by far the likeliest candidate for the dominant quenching mechanism given the SFH measured by \citet{Weisz2011}.  For our estimated infall time of 2--6 Gyr, we find that we need a time-averaged mass-loading factor of $\overline{\eta}=6-11$ to remove cold gas from DDO~113 with strangulation alone, compared to the literature values of $\eta \sim 3-90$ \citep{Bordoloi2014,Muratov2015,Lu2015,Christensen2016}. \par

This finding is surprising, as strangulation is typically thought to require the host to have a hot, virialized gas halo, though recent numerical results \citep[e.g.,][]{Birnboim2003,Keres2009} have shown that a minimum halo mass of only $\sim 2 \times 10^{11} \, M_{\odot}$ is required at $z=0$ for the formation of such a halo. DDO 113's host, NGC 4214, appears to lies slightly below this halo mass threshold. On the other hand, recent observational and theoretical work on low-mass hosts indicate the presence of large, outflow-driven cool gas reservoirs. We use a simulation-driven model for the cool CGM of NGC~4214 to show that, while ram pressure stripping of DDO 113 by the CGM likely plays a part in removing gas from DDO 113, its more important role is likely to enforce the dominant strangulation process by ensuring gas ejected from DDO 113 never returns. Our work shows, for the first time, the key role that the CGM may play in environmental quenching of satellites around sub-L\textsubscript{*} hosts.

\hspace{\linewidth}
\section*{Acknowledgements}
We would like to thank Chris Hirata, Adam Leroy, Cassi Lochhaas, and Amy Sardone for helpful discussions regarding quenching scenarios. C.G., C.S.K., and A.H.G.P. are supported by the NSF Grant No. AST-1615838. C.S.K. is also supported by NSF grants AST-1515876,
AST-1515927 and AST-1814440. Research by D.J.S. is supported by NSF grants AST-1821987, AST-1821967, AST-1813708, AST-1813466, and AST-1908972. Research by D.C. is supported by NSF grant AST-1814208, and by NASA through grants number HST-GO-15426.007-A and HST-GO-15332.004-A from the Space Telescope Science Institute, which is operated by AURA, Inc., under NASA contract NAS 5-26555.\par
The LBT is an international collaboration among institutions in the United States, Italy and Germany. LBT Corporation partners are: The University of Arizona on behalf of the Arizona university system; Istituto Nazionale di Astrofisica, Italy; LBT Beteiligungsgesellschaft, Germany, representing the Max-Planck Society, the Astrophysical Institute Potsdam, and Heidelberg University; The Ohio State University, and The Research Corporation, on behalf of The University of Notre Dame, University of Minnesota and University of Virginia.\par
This research has made use of the NASA/IPAC Extragalactic Database (NED), which is operated by the Jet Propulsion Laboratory, California Institute of Technology, under contract with the National Aeronautics and Space Administration. This research has made use of NASA's Astrophysics Data System.

\bibliographystyle{mnras}
\bibliography{library,annikarefs}
\bsp	
\label{lastpage}
\end{document}